\documentclass[twocolumn,preprintnumbers,amsmath,amssymb,showpacs]{revtex4}
\usepackage{epsfig}
\begin{document}
\setlength{\voffset}{1.0cm}
\title{Baryons in the large $N$ limit of the massive two-dimensional \\ Nambu--Jona-Lasinio model}
\preprint{FAU-TP3-08/03}
\author{Christian Boehmer}
\author{Felix Karbstein}
\author{Michael Thies\footnote{Electronic addresses: \\christian.boehmer,felix,thies@theorie3.physik.uni-erlangen.de}}
\affiliation{Institut f\"ur Theoretische Physik III,
Universit\"at Erlangen-N\"urnberg, D-91058 Erlangen, Germany}
\date{\today}
\begin{abstract}
Baryons in the massive Nambu--Jona-Lasinio model in 1+1 dimensions (the massive chiral Gross-Neveu model)
are studied in the limit of an infinite number of flavors. The baryon mass is evaluated for a wide range of bare fermion
masses and filling fractions, combining analytical asymptotic expansions with a full numerical Hartree-Fock
calculation.
\end{abstract}
\maketitle
\section{Introduction}\label{sect1}
Since the seminal paper by Gross and Neveu in 1974 \cite{L1}, field theoretic models of fermions in 1+1 dimensions
with (broken or unbroken) chiral symmetry have turned out to be quite useful. In view of a long list of applications
in particle, condensed matter and mathematical physics, it is surprising that after more than three decades the basic
solution of these models is still incomplete, even in the tractable limit of an infinite number of flavors $N$. A few years ago it has been
realized that the alleged phase diagram of Gross-Neveu (GN) models in the temperature ($T$) versus
chemical potential ($\mu$) plane \cite{L2,L3} was inconsistent with the known baryon spectrum. In the standard GN model 
with scalar-scalar interaction, this problem has been solved in the meantime with the construction of a soliton crystal phase
(see \cite{L4} and references therein) built out of Dashen-Hasslacher-Neveu baryons \cite{L5}.
However there are still large gaps in our understanding of the model with continuous chiral symmetry, i.e., the 
Nambu--Jona-Lasinio model \cite{L6} in two dimensions (NJL$_2$). In the chiral limit, it is known that solitonic baryons
and soliton crystals do exist here as well, although their properties are quite distinct from those of the standard GN model \cite{L7}.
In particular, induced fermion number plays an important role in the NJL$_2$ model \cite{L8}. By contrast, in the massive
NJL$_2$ model, very little is known yet about either baryons or the phase
diagram. We are aware of only two works addressing the issue of baryons at finite bare fermion mass. Salcedo et al. \cite{L9} use an
elegant variational ansatz to reduce the problem to the sine-Gordon kink, both in the 't~Hooft model \cite{L9a} and the NJL$_2$ model,
and point out the close relationship to the Skyrme picture in higher dimensions \cite{L9b}. 
In Ref.~\cite{L10} the sine-Gordon equation was subsequently identified as the first term of a systematic chiral expansion
and higher order corrections were determined with the help of derivative expansion techniques. In the present work, we
try to further improve this situation by constructing
baryons in the (large $N$) NJL$_2$ model for arbitrary bare fermion mass, using a combination of analytical
and numerical methods. Our ultimate goal is the phase diagram of the model which will be discussed elsewhere.
Like in the GN model, we expect the baryon mass to play the role of the critical chemical potential where a 
soliton crystal phase sets in at zero temperature. Aside from their importance for the
phase diagram, baryons in the massive NJL$_2$ model are also of interest in their own right. By varying the bare fermion
mass, we have the unique possibility to interpolate between the Skyrme picture characteristic for the chiral limit and
the non-relativistic valence picture expected to hold in the heavy fermion limit.

Let us now pose the baryon problem at large $N$ in a more precise fashion. The Lagrangian of the 
(massive) NJL$_2$ model reads
\begin{equation}
{\cal L}= \bar{\psi} ({\rm i}\partial \!\!\!/-m_0)\psi + \frac{g^2}{2}\left[ (\bar{\psi}\psi)^2+(\bar{\psi}{\rm i}\gamma_5\psi)^2\right]
\label{a1}
\end{equation}
where flavor indices are suppressed as usual, i.e., $\bar{\psi}\psi=\sum_{k=1}^N \bar{\psi}_k \psi_k$ etc.
In the large $N$ limit the Hartree-Fock (HF) approximation \cite{L11}, or, equivalently, the stationary phase approximation
to the functional integral \cite{L5}, become exact. The Dirac-HF equation 
\begin{equation}
\left(-\gamma_5 {\rm i}\partial_x + \gamma^0 S(x) + {\rm i}\gamma^1 P(x)\right)\psi_{\alpha}=E_{\alpha}\psi_{\alpha}
\label{a2}
\end{equation}
for the single particle orbital $\alpha$ has to be solved simultaneously with the 
self-consistency conditions
\begin{eqnarray}
S & = & m_0  -Ng^2 \sum_{\alpha}^{\rm occ} \bar{\psi}_{\alpha}\psi_{\alpha},
\nonumber \\
P & = & -Ng^2 \sum_{\alpha}^{\rm occ} \bar{\psi}_{\alpha}{\rm i}\gamma_5 \psi_{\alpha}.
\label{a3}
\end{eqnarray}
The HF energy
\begin{equation}
E_{\rm HF} = \sum_{\alpha}^{\rm occ} E_{\alpha} + \frac{1}{2g^2}\int {\rm d}x \left[ (S-m_0)^2+P^2 \right]
\label{a4}
\end{equation}
comprises the sum over single particle energies of all occupied states (including the Dirac sea) and the standard correction
for double counting of the potential energy.
The vacuum problem is a special case of Eqs.~(\ref{a2}-\ref{a4}) where $S=m$ (the physical fermion mass), $P=0$, and all negative
energy states are filled. Self-consistency then yields the gap equation (in units where $m=1$, and using the ultraviolet (UV) 
cutoff $\Lambda/2$ \cite{L10})
\begin{equation}
\frac{\pi}{Ng^2} = \gamma + \ln \Lambda.
\label{a5}
\end{equation}
Here we have introduced the ``confinement parameter" $\gamma$ \cite{L12,L12a},
\begin{equation}
\gamma = m_0 \ln \Lambda,
\label{a6}
\end{equation}
where the name (borrowed from condensed matter physics) refers to confinement of kink and antikink, not of the elementary
fermions.
Trading the bare parameters ($g^2,m_0$) for the physical parameters ($m=1,\gamma$) with the help of Eqs.~(\ref{a5},\ref{a6})
is all that is needed to renormalize the model and eliminate divergences in the baryon problem. 
Baryons in the HF approach are characterized by $x$-dependent potentials $S(x),P(x)$ and
one (partially or fully) occupied extra level as compared to the vacuum. The observables of most interest to us are
the baryon mass, defined as the difference in HF energies between baryon state and vacuum, the self-consistent
potentials $S,P$ and the fermion density.  

We shall use both analytical and numerical methods, depending on the parameters. The paper is organized accordingly. 
In Sec.~\ref{sect2}, we remind the reader of the derivative expansion and collect the results relevant for the vicinity of the
chiral 
limit. In Sec.~\ref{sect3} we apply a recently developed effective field theory for the valence level \cite{L13} to get
complementary analytical insight into the non-relativistic regime. The full numerical solution of the HF problem including the
Dirac sea is presented in Sec.~\ref{sect4}. Sec.~\ref{sect5} contains a short summary and our conclusions.
\section{Derivative expansion}\label{sect2}
In general the HF problem as defined through Eqs.~(\ref{a2},\ref{a3}) is rather involved. This raises the question whether
one can bypass its full solution, at least in some regions of parameter space.
Near the chiral limit, the potentials $S,P$ become very smooth. It is then possible to ``integrate out" the fermions approximately, 
using the derivative expansion technique, see e.g. Refs.~\cite{L14,L15}. This method presupposes full occupation
of each level with $N$ fermions and is inapplicable for partially occupied levels. It results in
a purely bosonic effective field theory
for the complex scalar field $\Phi=S-{\rm i}P$ which, in the large $N$ limit, can be treated classically. 
Hence, one gets direct access to the HF potential without need to solve the Dirac-HF equation self-consistently.
The leading order reproduces the sine-Gordon approach of Ref.~\cite{L9}.
In Refs.~\cite{L10,L16} this program was carried through to rather high order in the derivative expansion, yielding a
systematic expansion in the ratio of pion mass to physical fermion mass. The field $\Phi$  
was computed in polar coordinates,
\begin{equation}
\Phi = (1+\lambda){\rm e}^{2{\rm i}\chi},
\label{b1}
\end{equation}
with $\lambda,\chi$ expressed in terms of the pion mass $m_{\pi}$ and the spatial variable $\xi=m_{\pi} x$.
For the present purpose, we express everything in terms of the confinement parameter $\gamma$, using the 
(approximate) relationship
\begin{equation}
m_{\pi} = 2 \sqrt{\gamma} \left(1- \frac{1}{3}\gamma + \frac{11}{90} \gamma^2 - \frac{5}{126}\gamma^3 \right).
\label{b2}
\end{equation}
To this order in $\gamma$, the baryon mass is given by
\begin{equation}
M_{\rm B} = \frac{4\sqrt{\gamma}N}{\pi} \left( 1  - \frac{4}{9}\gamma + \frac{9}{50}\gamma^2- \frac{101}{735} \gamma^3 \right),
\label{b3}
\end{equation}
whereas the result for the potentials translates into 
\begin{eqnarray}
S & = & 1 - \frac{2}{\cosh^2 \xi} + \frac{2}{\cosh^4 \xi}\gamma 
\nonumber \\
& - &  \frac{2}{9}\left( \frac{20}{\cosh^2\xi}
-\frac{55}{\cosh^4\xi}+\frac{41}{\cosh^6\xi}\right) \gamma^2,
\nonumber \\
P & = & \frac{2 \sinh \xi}{\cosh^2 \xi} \left[ 1  - \frac{1}{2} \left(1 + \frac{2}{\cosh^2 \xi}\right) \gamma \right. 
\nonumber \\
& + &  \left. \frac{1}{72}\left( 47 - \frac{276}{\cosh^2 \xi}+ \frac{328}{\cosh^4 \xi}\right) \gamma^2 \right].
\label{b4}
\end{eqnarray}
In the vicinity of the chiral limit, fermion number is induced by the topologically non-trivial
field $\Phi$ and thus only resides in the negative energy states. For the fermion number divided by $N$ (``baryon number"), 
the derivative expansion yields 
\begin{equation}
B=\int {\rm d}x \left(\frac{\chi'}{\pi}\right)= \frac{1}{\pi}\left[ \chi(\infty)-\chi(-\infty)\right],
\label{b5}
\end{equation}
relating baryon number to chiral U(1) winding number of the field $\Phi$, much like in the Skyrme model. One cannot conclude from
this result that the induced fermion density is given by $\chi'/\pi$ though.
As discussed in Ref.~\cite{L8}, the fermion density can be determined 
from the equation for the partially conserved axial current (PCAC),
\begin{equation}
\partial_{\mu}j_5^{\mu} =2m_0 \bar{\psi}{\rm i}\gamma_5 \psi.
\label{b6}
\end{equation}
Taking the expectation value of Eq.~(\ref{b6}) in the baryon state, using the relationship 
$j_5^{\mu}=\epsilon^{\mu \nu}j_\nu$ (valid in two dimensions) and integrating over $x$,
one finds the fermion density per flavor
\begin{equation}
\rho = - \frac{2 \gamma}{\pi} \int_{-\infty}^x {\rm d}x' P(x').
\label{b7}
\end{equation}
An expansion in $\gamma$ then yields in the present case
\begin{eqnarray}
\rho &=& \frac{2 \sqrt{\gamma}}{\pi \cosh \xi}\left[1-\frac{1}{6}\left(1+\frac{2}{\cosh^2 \xi}\right)\gamma \right.
\nonumber \\
&+ &  \left. \frac{1}{360}\left(171- \frac{500}{\cosh^2 \xi}+ \frac{328}{\cosh^4 \xi}\right)\gamma^2 \right].
\label{b8}
\end{eqnarray}
Judging from the convergence properties of the series in $\gamma$, we expect expressions (\ref{b2}-\ref{b8}) to be quantitatively reliable
up to $\gamma \approx 0.2$. 

Let us use these results to exhibit some differences between baryons in massive GN and NJL$_2$ models,
respectively. First of all, in the GN model, there is no induced, but only valence fermion density. 
The negative energy states cancel exactly when taking the difference between the fermion density of the baryon and the vacuum \cite{L10}.  
Turning to the NJL$_2$ model, the situation is different. 
\begin{figure}
\begin{center}
\epsfig{file=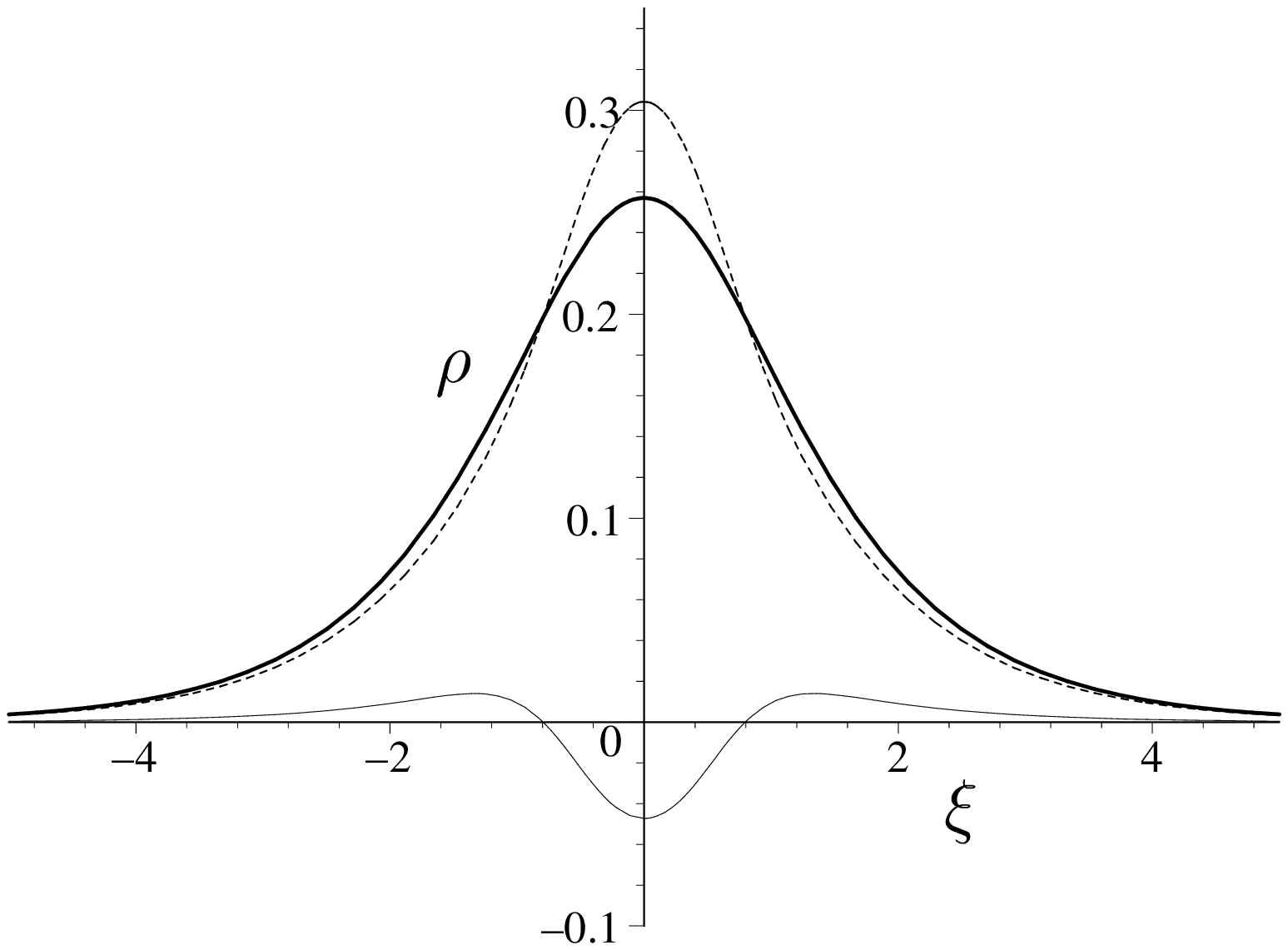,angle=270,width=8cm}
\caption{Thick solid line: induced fermion density (\ref{b8}), dashed line: leading order term $\chi'/\pi$, thin line: difference, according to 
derivative expansion ($\gamma=0.2$).}
\label{fig1}
\end{center}
\end{figure}
In Fig.~\ref{fig1} we illustrate the induced fermion density for the case
$\gamma=0.2$ and compare it
with the naive expectation $\chi'/\pi$. The difference integrates to 0, in agreement with Eq.~(\ref{b5}).
Another pertinent observation is the following. In the GN model, the scalar potential $S$ 
is reflectionless for any static solution \cite{L4,L17}. This has turned out to be instrumental for the exact, analytical solvability of the
baryon problem.  
We can now easily check whether the same is true for the NJL$_2$ model, at least in the region of validity of the derivative expansion. 
\begin{figure}
\begin{center}
\epsfig{file=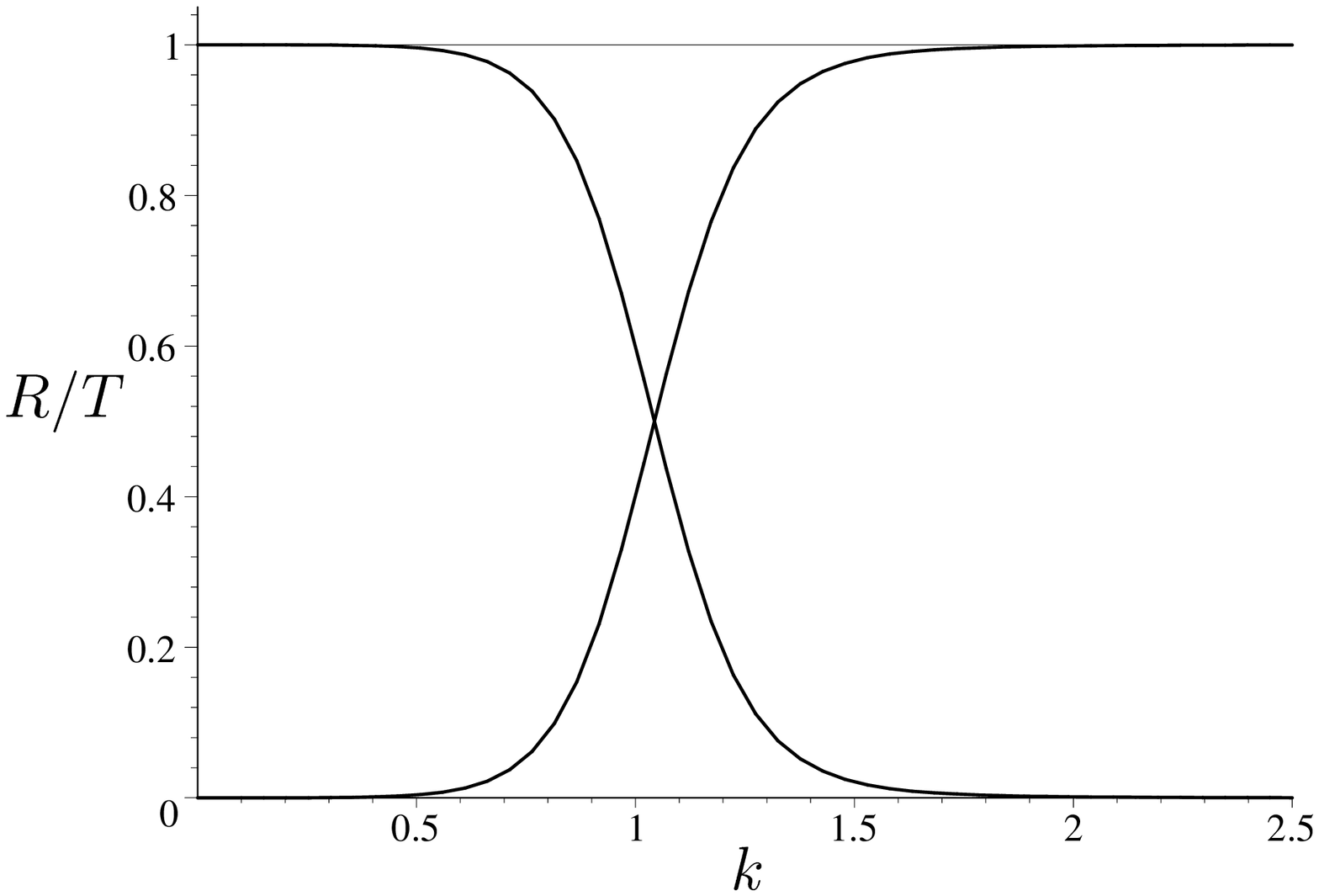,angle=270,width=8cm}
\caption{Reflection coefficient $R$ (falling curve) and transmission coefficient $T$ (rising curve) for baryon HF potential (\ref{b4}) 
versus momentum ($\gamma=0.2$).}
\label{fig2}
\end{center}
\end{figure}
By solving the Dirac scattering problem with potential (\ref{b4}), we find that this potential is not transparent, 
see Fig.~\ref{fig2} for the case $\gamma=0.2$. The transmission coefficient rises from 0 at low energies to 1 at high energies. Similar results are
obtained at other values of $\gamma$. This is at variance with claims in the literature \cite{L18} (see however Ref.~\cite{L19}) and already
indicates that the baryon problem in the NJL$_2$ model is more challenging than in the GN model.
\section{No-sea effective theory}\label{sect3} 
Recently, an approximation has been devised which yields analytical insight into 
the regime where the baryon features one positive energy valence level, while still having smooth HF potentials.
The idea is to integrate out the negative energy fermions (i.e., the Dirac sea) and derive an effective theory for
the (positive energy) valence fermions only. The HF equation for the baryon then reduces to a single non-linear (and in general non-local)
Dirac equation. This is a tremendous simplification as compared to the full relativistic HF problem, although 
somewhat more involved than the derivative expansion of Sec.~\ref{sect2}, where the fermions have been integrated out
altogether. For the derivation of the effective Lagrangian in the massive NJL$_2$ model, we refer to Ref.~\cite{L13}.
Here we concentrate on the solution of the non-linear Dirac equation and the determination of the parameter region
where the proposed truncation is valid.

We proceed as follows. The effective no-sea Lagrangian consists of a massive free Dirac part and interaction
terms descending from the original scalar and pseudoscalar four-fermion interactions. 
As far as terms involving only scalar interactions are concerned, we keep all contributions in Eq.~(39) of Ref. \cite{L13}.
The terms involving pseudoscalar interactions require one additional approximation. Starting point is the result for
the massive NJL$_2$ model, Eq.~(76) of \cite{L13}, replacing the scalar effective coupling constant $g_{\rm eff}^2=\pi/N$
by the corresponding expression in the massive model (cf. Eq.~(72) of \cite{L13}). 
If we insist on an analytical solution as we do in this section, we are only able to handle the case in which the
non-locality due to one-pion exchange is of short range. We therefore assume that the mass term $m_{\pi}^2=4\gamma$
dominates the inverse pion propagator and expand in the remaining two terms as follows,
\begin{eqnarray}
 & & \!\!\!\!\frac{2\pi}{N} \bar{\psi}{\rm i}\gamma_5\psi \frac{1}{\square + m_{\pi}^2-\frac{4\pi}{N(1+\gamma)}
\bar{\psi}\psi}\bar{\psi}{\rm i}\gamma_5\psi \ \approx
\label{c1} \\
& &  \frac{2\pi}{N}\bar{\psi}{\rm i}\gamma_5\psi
\frac{1}{m_{\pi}^2}\left(1- \frac{\square}{m_{\pi}^2}+
\frac{4\pi \bar{\psi}\psi}{N m_{\pi}^2(1+\gamma)}\right) \bar{\psi}{\rm i}\gamma_5\psi.
\nonumber
\end{eqnarray}
The result is added to the GN model effective Lagrangian. At this stage it is difficult to predict the precise range
of applicability of the approximation due to the unavoidable issue of self-consistency. We will return to this question
once we have solved the non-linear Dirac-HF equation. The final effective Lagrangian reads ($m=1$)
\begin{widetext}
\begin{eqnarray}
{\cal L}_{\rm eff} & = & \bar{\psi}({\rm i}\partial \!\!\!/ - 1)\psi + \frac{\pi}{2N}\frac{1}{(1+\gamma)}(\bar{\psi}\psi)^2
 - \frac{\pi}{24 N}\frac{1}{(1+\gamma)^2}(\square \bar{\psi}\psi)(\bar{\psi}\psi)
 + \frac{\pi^2}{6N^2}\frac{1}{(1+\gamma)^3}(\bar{\psi}\psi)^3 
\nonumber \\
& + &  \frac{11\pi}{1440N}\frac{(1+6\gamma/11)}{(1+\gamma)^3}(\square^2 \bar{\psi}\psi)(\bar{\psi}\psi)
 - \frac{\pi^2}{12 N^2}\frac{(1+\gamma/2)}{(1+\gamma)^4}(\square \bar{\psi}\psi)(\bar{\psi}\psi)^2
+ \frac{\pi^3}{6N^3}\frac{(1+\gamma/4)}{(1+\gamma)^5}(\bar{\psi}\psi)^4 
\nonumber \\
& + &  \frac{\pi}{2N\gamma}(\bar{\psi}{\rm i}\gamma_5\psi)^2
 - \frac{\pi}{8N\gamma^2} (\square \bar{\psi}{\rm i}\gamma_5 \psi)
 (\bar{\psi}{\rm i}\gamma_5 \psi)
 + \frac{\pi^2}{2N^2 \gamma^2}\frac{1}{(1+\gamma)}(\bar{\psi}{\rm i}\gamma_5\psi)^2(\bar{\psi}\psi).
\label{c2}
\end{eqnarray}
\end{widetext}
By construction, it is to be used in the positive energy sector only, since the effects of the negative energy states
are already encoded
in the Lagrangian. For a single baryon, the ensuing Euler-Lagrange equation is the non-linear Dirac equation for the (normalized) valence spinor,
\begin{equation}
\left( - \gamma_5 {\rm i} \partial_x + \gamma^0 S + {\rm i}\gamma^1 P\right)\psi_0=E_0 \psi_0, \quad  \int{\rm d}x\, \psi_0^{\dagger}
\psi_0=1.
\label{c3}
\end{equation}
Assuming the valence level to be occupied with filling fraction $\nu=n/N$ (where $n$ is the valence fermion number)
and using $\eta=\pi \nu$ for ease of notation (all $\pi$'s disappear),
the scalar and pseudoscalar potentials are expressed self-consistently as ($'=\partial_x$)
\begin{eqnarray}
S&=& 1-\frac{\eta s_0}{(1+\gamma)} - \frac{\eta s_0''}{12(1+\gamma)^2} -\frac{\eta^2 s_0^2}{2(1+\gamma)^3}
\nonumber \\
& -  &  \frac{11\eta (1+6\gamma/11) s_0^{IV}}{720 (1+\gamma)^3}
 -\frac{\eta^2(1+\gamma/2)}{6(1+\gamma)^4}\left(2 s_0''s_0+(s_0')^2\right)
\nonumber \\
& - &  \frac{2 \eta^3 (1+ \gamma/4)s_0^3}{3(1+\gamma)^5}- \frac{\eta^2 p_0^2}{2\gamma^2(1+\gamma)},
\nonumber \\
P&=&-\frac{\eta p_0}{\gamma} - \frac{\eta p_0''}{4\gamma^2}- \frac{\eta^2 p_0s_0}{\gamma^2(1+\gamma)}
\label{c4}
\end{eqnarray}
through scalar ($s_0$) and pseudoscalar ($p_0$) valence level condensates
\begin{equation}
s_0 = \bar{\psi}_0 \psi_0, \quad p_0 = \bar{\psi}_0 {\rm i}\gamma_5 \psi_0.
\label{c5}
\end{equation}
It seems hopeless to solve the complicated non-linear Dirac equation (\ref{c3}-\ref{c5}) in closed analytical form.
However, since the Lagrangian (\ref{c2}) is an approximate one, it is sufficient to solve the equation
perturbatively in some formal expansion parameter. 
We choose to expand in $\nu$ and postpone till later the discussion of the validity of the truncation scheme.
In order to solve the Dirac equation, we then find that we first have to solve a simple non-linear differential equation,
followed by a sequence of inhomogeneous, linear differential equations. The lengthy analytical computations can be
done conveniently with computer algebra (we used MAPLE), therefore we skip the details and record only the final results. 
The baryon mass  (computed from the classical field energy) has the following power series expansion in $\eta=\pi \nu$, 
\begin{eqnarray}
\pi \frac{M_{\rm B}}{N}  & =&   \eta - \frac{\eta^3}{24(1+\gamma)^2} + \frac{(\gamma^2-15\gamma-8)\eta^5}{1920 \gamma(1+\gamma)^5}
\nonumber \\
& - &  \frac{(\gamma^4-302 \gamma^3+265 \gamma^2+376 \gamma+88)\eta^7}{322560 \gamma^2(1+\gamma)^8}.
\label{c6}
\end{eqnarray}
The energy $E_0$ of the valence level is closely related to $M_{\rm B}$. According to standard HF theory,
the single particle energy can be interpreted as removal energy of a fermion, or equivalently
\begin{equation}
E_0=\frac{\partial M_{\rm B}}{\partial n}. 
\label{c7}
\end{equation}
This is indeed what we find in the calculation, so that there is no need to spell out $E_0$. 
Next we turn to the HF potentials (\ref{c4}) derived from the self-consistent valence level spinor $\psi_0$.
For the scalar and pseudoscalar potentials, we obtain
\begin{eqnarray}
S & = & 1+ \frac{s_{22}}{\cosh^2 \xi}\eta^2 + \left( \frac{s_{42}}{\cosh^2 \xi}+\frac{s_{44}}{\cosh^4\xi}\right)\eta^4
\nonumber \\
& + &  \left( \frac{s_{62}}{\cosh^2 \xi}+\frac{s_{64}}{\cosh^4\xi} + \frac{s_{66}}{\cosh^6 \xi}\right)\eta^6
\label{c8} \\
P & = & \left[ \frac{p_{33}}{\cosh^3 \xi}\eta^3 +  \left(\frac{p_{53}}{\cosh^3 \xi}+ \frac{p_{55}}{\cosh^5 \xi}\right)\eta^5\right] \sinh \xi
\nonumber
\end{eqnarray}
with $\gamma$-dependent coefficients $s_{nm},p_{nm}$ relegated to the appendix in order to keep the paper readable
(the subscripts $n,m$ denote the powers of $\eta$ and $1/\cosh \xi$, respectively).
The spatial variable $\xi=yx$ in Eq.~(\ref{c8}) involves a scale factor
\begin{eqnarray}
y  &=&  \frac{\eta}{2(1+\gamma)} -\frac{(\gamma^2-3\gamma-2)\eta^3}{48 \gamma(1+\gamma)^4}
\nonumber \\
& +  &  \frac{(3 \gamma^4-226 \gamma^3-65 \gamma^2+ 68 \gamma + 24)\eta^5}{11520 \gamma^2(1+\gamma)^7}.
\label{c12}
\end{eqnarray}
One verifies that it is related to the single particle energy $E_0$ via 
\begin{equation}
E_0=\sqrt{1-y^2}.
\label{c13}
\end{equation}
This last relation can be understood in physics terms as follows. Asymptotically, the valence wave function falls off exponentially with the 
$\kappa$ value (imaginary momentum) of the bound state. Due to self-consistency, the same parameter will govern the
asymptotic exponential decay of the potentials in a no-sea HF calculation. This is exactly what Eq.~(\ref{c13}),
together with the shape of the potentials $S$ and $P$, guarantees.

We note in passing that the asymptotic expansions for the scalar potential $S$ at small $\gamma$, Eq.~(\ref{b4}), and at large $\gamma$
or small $\nu$, Eq.~(\ref{c8}), have the same functional form. This is not true for the pseudoscalar potential $P$ where Eq.~(\ref{b4})
involves even, Eq.~(\ref{c8}) odd powers of $1/\cosh \xi$.   
 
We have tested the above results in two different ways. If we switch off the pseudoscalar coupling, we can carry out the same calculation for
the massive GN model and reproduce the well-known exact results to the expected accuracy. 
A more specific test of the NJL$_2$ model calculation is provided by the divergence of the axial current implying \cite{L8}
\begin{equation}
\partial_x \rho = - \frac{2\gamma}{\pi} P.
\label{c14}
\end{equation}
Here $\rho$ denotes the fermion density per flavor
consisting of the valence density and the induced part from the Dirac sea ($n=\nu N$),
\begin{equation}
\rho = \nu \psi_0^{\dagger}\psi_0 - \frac{1}{2\pi} \partial_x P.
\label{c15}
\end{equation}
Eqs.~(\ref{c14},\ref{c15}) yield the following non-trivial identity relating the valence fermion density to the pseudoscalar HF potential,
\begin{equation}
\nu \partial_x (\psi_0^{\dagger}\psi_0) = -\frac{2 \gamma}{\pi}P + \frac{1}{2\pi} \partial_x^2 P .
\label{c16}
\end{equation} 
In our calculation it is violated at order $\nu^7$. For the sake of completeness, let us quote the separate results 
for the valence fermion density $\rho_{\rm val}=\nu \psi_0^{\dagger}\psi_0$,
\begin{eqnarray}
\pi \rho_{\rm val} & = &   \frac{v_{22}}{\cosh^2 \xi}\eta^2 + \left( \frac{v_{42}}{\cosh^2 \xi}+\frac{v_{44}}{\cosh^4\xi}\right)\eta^4
\nonumber \\
& + &  \left( \frac{v_{62}}{\cosh^2 \xi}+\frac{v_{64}}{\cosh^4\xi} + \frac{v_{66}}{\cosh^6 \xi}\right)\eta^6,
\label{c17}
\end{eqnarray}
and for the induced fermion density $\rho_{\rm ind}=-\frac{1}{2\pi}\partial_x P$,
\begin{eqnarray}
\pi \rho_{\rm ind} & = &  \left( \frac{i_{42}}{\cosh^2 \xi}+\frac{i_{44}}{\cosh^4\xi}\right)\eta^4
\nonumber \\
& + &  \left( \frac{i_{62}}{\cosh^2 \xi}+\frac{i_{64}}{\cosh^4\xi} + \frac{i_{66}}{\cosh^6 \xi}\right)\eta^6,
\label{c18} 
\end{eqnarray}
where the coefficients can again be found in the appendix.
 
Let us now discuss the range of validity of our truncation. The most obvious candidate is the regime 
\begin{equation}
\gamma \gg 1 \qquad {\rm (regime\ I)}
\label{c19}
\end{equation}
and arbitrary $\nu$.
In this case we can read off from our results that $\bar{\psi}\psi \sim 1/\gamma$, $\bar{\psi}{\rm i}\gamma_5\psi
\sim 1/\gamma^2$ and $\square \sim 1/\gamma^2$. Inspection of the coefficients in ${\cal L}_{\rm eff}$ then shows that 
we have kept all terms up to order $1/\gamma^8$ in an asymptotic expansion for large $\gamma$. Since the identity (\ref{c16})
is violated at order $1/\gamma^7$ we can trust the results (expanded in $1/\gamma$) up to order $1/\gamma^6$.
By way of example, we note for future reference the asymptotic behavior of $M_{\rm B}$ for $\gamma \to \infty$ and full occupation of the 
valence level ($\nu=1$) complementary to Eq.~(\ref{b3}),
\begin{eqnarray}
\frac{M_{\rm B}}{N} &=& 1- \frac{\pi^2}{24 \gamma^2} + \frac{\pi^2}{12 \gamma^3} - \frac{\pi^2(240-\pi^2)}{1920 \gamma^4}
\label{c20} \\
& + & \frac{\pi^2(16-\pi^2)}{96 \gamma^5} - \frac{\pi^2 (67200-13776 \pi^2+ \pi^4)}{322560 \gamma^6}.
\nonumber
\end{eqnarray}
Another case of interest is weak occupation of the valence level, i.e., taking the formal expansion in $\nu$
literally. Here, $\bar{\psi}\psi \sim \nu^2, \bar{\psi}{\rm i}\gamma_5 \psi \sim \nu^3, \square \sim \nu^2$ and
all terms up to order $\nu^8$ are kept in ${\cal L}_{\rm eff}$. A necessary condition for this to be valid is $\nu \ll 1$.
Due to the expansion of the propagator in Eq.~(\ref{c1}), inverse powers of $\gamma$ appear in the 
coefficients, and we have to require in addition $\nu^2 \ll \gamma$. In this regime, all terms are then consistently kept
and, according to the test based on Eq.~(\ref{c16}), the results are trustworthy up to order $\nu^6$. This 2nd regime of
applicability may be summarized as
\begin{equation}
\nu \ll {\rm min} (1,\sqrt{\gamma}\,) \qquad {(\rm regime\ II)}.
\label{c21}
\end{equation}
If the condition $\nu \ll \sqrt{\gamma}$ does not hold, the non-locality of the effective action becomes long range and
our approximation breaks down. 

We wind up this section with a summary of the analytical results obtained in Secs.~\ref{sect2} and \ref{sect3}. From a practical point of view,
it is fairly easy to judge the quality of an effective theory by comparing successive orders in the expansion.
\begin{figure}
\begin{center}
\epsfig{file=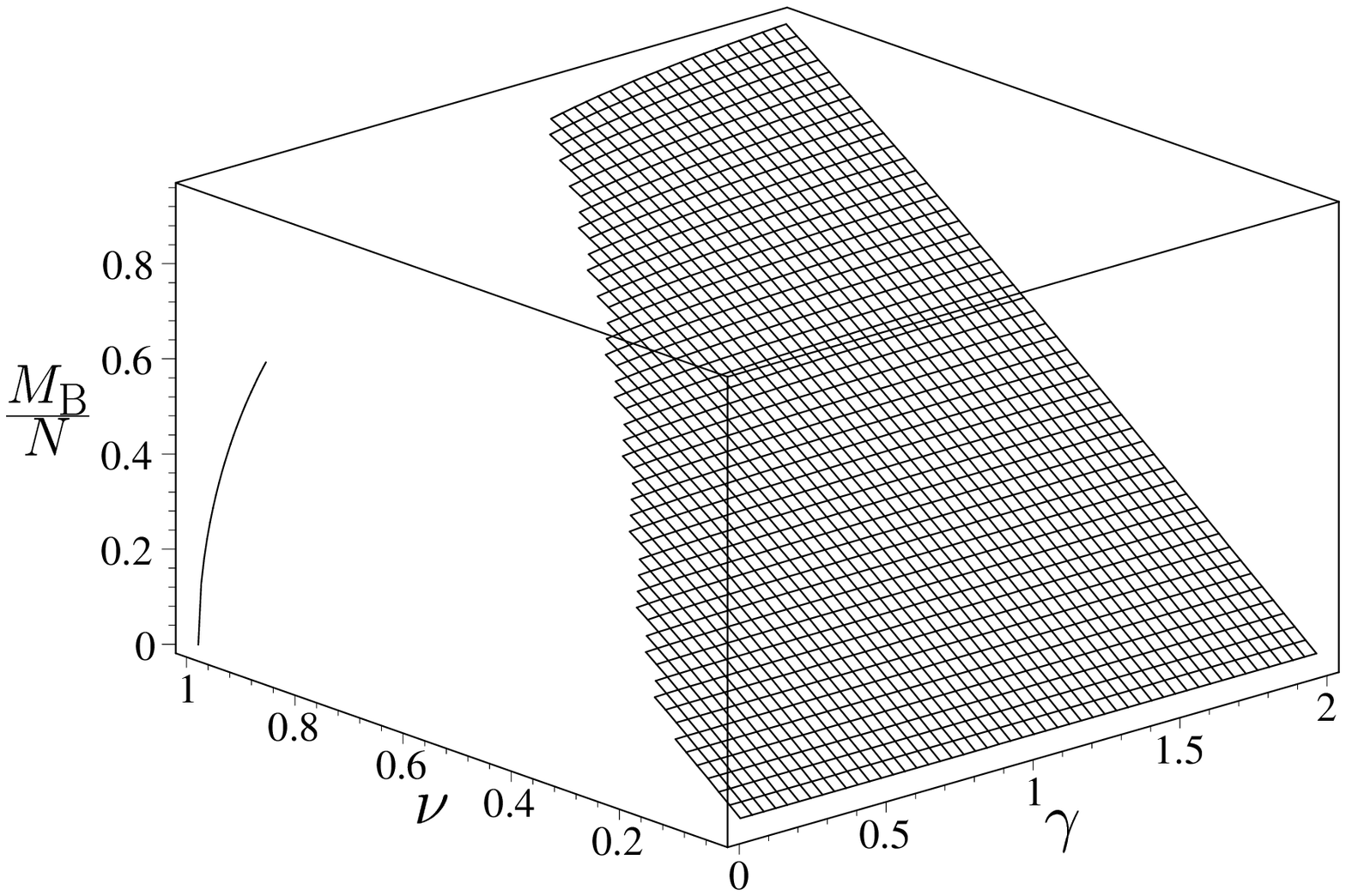,height=8cm,width=6.4cm,angle=270}
\caption{Analytical results for the baryon mass as a function of filling fraction $\nu$ and confinement parameter $\gamma$.
Shown are only results which are stable at the 0.2$\%$ level if one removes the highest order correction term. Curve
starting at $\gamma=0,\nu=1$: derivative expansion. Tilted surface: No-sea effective theory.}
\label{fig3}
\end{center}
\end{figure}
This is illustrated in Fig.~\ref{fig3} where we plot the baryon mass (per flavor) $M_{\rm B}/N$ as computed from Eqs.~(\ref{b3}) and (\ref{c6})
versus $\nu$ and $\gamma$. 
We only display the result if the highest order correction kept 
contributes less than 0.002 of the total result (other numerical values for the tolerance would give qualitatively similar plots). This 
shows clearly
that the derivative expansion (curve starting at $\nu=1,
\gamma=0$) and the no-sea effective theory (tilted surface) are valid in disjoint regions of the diagram, separated by a gap. Needless to
 say, the  
no-sea calculation is also valid at higher values of $\gamma$ not shown here. The structure of the baryons is 
very different, depending on which asymptotic expansion we look at. Baryons described by the derivative expansion have 
no filled positive energy level, so that baryon number is entirely due to fermion number induced in the Dirac sea.
If on the other hand the no-sea effective theory applies, there has to be a positive energy valence level, and the valence
fermion density dominates over the induced one, cf. the powers of $\eta$ in Eqs.~(\ref{c17},\ref{c18}). Our formulae also show that 
in the limit of large $\gamma$ for fixed $\nu$, or in the limit of small $\nu$ for fixed $\gamma$, the pseudoscalar terms
are suppressed and the NJL$_2$ results converge to those of the GN model. In fact, a non-relativistic
approximation to the Dirac equation would be perfectly adequate in this regime.
The region where the uppermost filled level reaches the middle of the gap and crosses zero energy
lies in the white part of Fig.~\ref{fig3} and has to await a numerical treatment, to which we now turn.
\section{Numerical Dirac-Hartree-Fock calculation}\label{sect4}
\subsection{Hamiltonian in discretized momentum space}\label{sect4a}
We first set up the Dirac Hamiltonian in the form of a matrix which can readily be diagonalized numerically.
Consider the Hamiltonian
\begin{equation}
H=\gamma_5 \frac{1}{\rm i} \frac{\partial}{\partial x} + \gamma^0 S(x) + {\rm i}\gamma^1 P(x)
\label{d1}
\end{equation}
with given scalar and pseudoscalar potentials. We choose the representation 
\begin{equation}
\gamma^0 =  \sigma_1,\quad \gamma^1 = - {\rm i}\sigma_2, \quad \gamma_5 = \gamma^0 \gamma^1 =  \sigma_3
\label{d2}
\end{equation}
of the $\gamma$ matrices leading to a real symmetric Hamiltonian matrix below.
In order to discretize the spectrum of $H$, we work in a finite interval of length $L$ and impose anti-periodic boundary
conditions for fermions, resulting in the discrete momenta
\begin{equation}
k_n=\frac{2\pi}{L}\left(n+\frac{1}{2}\right) \quad (n\in \mathbb{Z}).
\label{d3}
\end{equation}
We decompose the potentials into Fourier series,
\begin{equation}
S(x)=\sum_{\ell} S_{\ell} {\rm e}^{{\rm i}2\pi \ell x/L}, \quad P(x)= {\rm i} \sum_{\ell} P_{\ell} {\rm e}^{{\rm i}2\pi \ell x/L}.
\label{d4}
\end{equation}
Guided by the analytical results at small and large $\gamma$ of the preceding sections, we assume that parity is not
spontaneously broken in the baryon. A proper choice of the origin in coordinate space then leads to the following
symmetry relations for the scalar and pseudoscalar potentials,  
\begin{equation}
S(x)=S(-x), \qquad P(x)=-P(-x).
\label{d5}
\end{equation}
Together with the reality conditions for $S$ and $P$, we then find that the coefficients $S_{\ell},P_{\ell}$ are real and satisfy
\begin{equation}
S_{-\ell}=S_{\ell}, \qquad P_{-\ell}=-P_{\ell},\qquad P_0=0.
\label{d6}
\end{equation}
The zero mode $S_0$ of $S(x)$ acts like a mass term and will be denoted by $m$ (not to be confused with the physical
fermion mass in the vacuum set equal to 1 throughout this paper). We decompose the Hamiltonian as follows,
\begin{eqnarray}
H & = & H_0 + V, \quad H_0 \ =\  \gamma_5 \frac{1}{\rm i} \frac{\partial}{\partial x} + \gamma^0 m, 
\nonumber \\
V &  = &  \gamma^0 (S(x)-m) + {\rm i}\gamma^1 P(x),
\label{d7}
\end{eqnarray}
and diagonalize the free, massive Hamiltonian $H_0$ first,
\begin{equation}
H_0|\eta,n \rangle = \eta \sqrt{k_n^2+m^2} |\eta,n \rangle.
\label{d8}
\end{equation}
Here, $\eta=\pm 1$ is the sign of the energy and $k_n$ the discrete momentum. The free eigenspinors are given by
\begin{equation}
\langle x | \eta,n \rangle = \frac{1}{\sqrt{2LE}} \left( \begin{array}{c} \eta \sqrt{E+\eta k}\\ \sqrt{E-\eta k} \end{array} \right)
{\rm e}^{{\rm i}k x}
\label{d9}
\end{equation}
with the shorthand notation $k=k_n,E=\sqrt{k^2+m^2}$. Evaluating matrix elements of $V$ in this basis, we find
\begin{equation}
\langle \eta',n'|V|\eta,n \rangle  =  \frac{1}{2\sqrt{EE'}} \sum_{\ell \neq 0}\delta_{n-n',\ell}\left( A_{\ell}^{(+)}S_{\ell}+A_{\ell}^{(-)}P_{\ell}\right)
\label{d10}
\end{equation}
($k'=k_{n'},E'=\sqrt{(k')^2+m^2}$) with
\begin{equation}
A_{\ell}^{(\pm)} =  \eta' \sqrt{E'+\eta'k'}\sqrt{E-\eta k}  \pm  \eta \sqrt{E+\eta k}\sqrt{E'-\eta'k'}.
\label{d11}
\end{equation}
As $H_0$ is diagonal, we can now easily construct the full matrix $\langle \eta',n'|H|\eta,n \rangle$. 
We truncate the basis at some momentum index $\bar{N}$, keep positive and negative energy
states with labels $-\bar{N}-1,...,\bar{N}$ and diagonalize the resulting $(4\bar{N}+4)\times (4\bar{N}+4)$-dimensional,
real symmetric matrix $H$ numerically.
\subsection{Perturbation theory deep down in the Dirac sea}\label{sect4b}
It is possible and in fact necessary to treat the negative energy states deep down in the Dirac sea perturbatively. First, this restricts
the Hamiltonian matrix to be diagonalized numerically to manageable size. Secondly, there is a logarithmic
UV divergence in the sum over negative energy states which gets cancelled by a similar divergence in the double counting correction.
This cancellation has to be accomplished analytically before we can hope to reliably extract the finite part.  
We therefore compute the eigenvalues of $H$ in 2nd order perturbation theory in $V$, in the large volume limit $L\to  \infty$. 
For the negative energy continuum states, we find
\begin{equation}
E_{\rm pert} = - E(k) - \sum_{\ell =1}^{\ell_{\rm max}}\frac{E(k)^2S_{\ell}^2 +k^2 P_{\ell}^2-2 k_{\ell}E(k)S_{\ell}P_{\ell}}
{(k^2-k_{\ell}^2) E(k)}
\label{d12}
\end{equation}
with $E(k)=\sqrt{k^2+m^2}$ and $k_{\ell}=\pi \ell/L$. This result is valid for $k\gg \pi \ell_{\rm max}/L$, where the inverse
denominators don't blow up.
By comparing the perturbative eigenvalues with the ones from diagonalization, we can check whether we
have taken into account a sufficient number of basis states and identify any eigenvalues affected by the 
truncation. The result is an optimal energy $E_{\min}$, the energy of the lowest single
particle level in the Dirac sea computed by matrix diagonalization.
\subsection{Computation of the baryon mass}\label{sect4c}
The baryon mass has to be evaluated relative to the vacuum, i.e., as a difference of the HF energies (\ref{a4})
of the baryon and the vacuum. In the NJL$_2$ model (like in the standard GN model, for that matter)
the self-consistent potentials in the vacuum are 
$S(x)=1,P(x)=0$ in appropriate units. The gap equation (\ref{a5}) serves to 
eliminate the bare coupling constant from the double counting correction in Eq.~(\ref{a4}).
For the present discussion it is helpful to split the calculation of the baryon mass into four distinct pieces
\begin{equation}
M_{\rm B}/N= \Delta E_1+ \Delta E_2 + \Delta E_3 + \Delta E_4
\label{d13}
\end{equation}
defined as follows. We first evaluate the sum over single particle energies for occupied states resulting from matrix
diagonalization (subtracting the vacuum),
\begin{equation}
\Delta E_1 = \sum_{E_{\rm min}}^{E_{\rm max}} \left( E_{\alpha}- E_{\alpha}^{\rm vac}\right).
\label{d14}
\end{equation}
$E_{\rm min}$ has been introduced before, $E_{\rm max}$ denotes the energy of the highest occupied level in the baryon.
The vacuum single particle energies $E_{\alpha}^{\rm vac}$ are computed in a finite interval (length $L$) from the free massive theory
 with $m=1$.
Next we decompose the perturbative single particle energies (\ref{d12}) according to their UV behavior,
\begin{eqnarray}
E_{\rm pert} &=& - E(k) - \sum_{\ell=1}^{\ell_{\rm max}}\frac{S_{\ell}^2+P_{\ell}^2}{E(k)}
\label{d15} \\
&-  &  \sum_{\ell =1}^{\ell_{\rm max}}\frac{E(k_{\ell})^2S_{\ell}^2 +k_{\ell}^2 P_{\ell}^2-2 k_{\ell}E(k)S_{\ell}P_{\ell}}
{(k^2-k_{\ell}^2) E(k)}.
\nonumber
\end{eqnarray}
The last term on the right-hand side behaves asymptotically like $1/k^2$ and can be integrated over momenta without cutoff. The resulting
 convergent 
integral defines the contribution $\Delta E_2$ to the baryon mass,
\begin{equation}
\Delta E_2\! =\! - 2L \! \int_{k_{\rm min}}^{\infty} \!\!\! \frac{{\rm d}k}{2\pi}\! \sum_{\ell =1}^{\ell_{\rm max}}\frac{E(k_{\ell})^2S_{\ell}^2 
+k_{\ell}^2 P_{\ell}^2-2 k_{\ell}E(k)S_{\ell}P_{\ell}}{(k^2-k_{\ell}^2) E(k)}
\label{d16}
\end{equation}
and can be computed analytically ($k_{\rm min}$ is determined by $E_{\rm min}$ introduced above).
The first two terms in Eq.~(\ref{d15}), after vacuum subtraction, give rise to an elementary, logarithmically 
divergent integral which has to be regularized with the same cutoff as the gap equation 
and will be denoted by $\Delta E_3$,
\begin{eqnarray}
\Delta E_3 &=& - 2 L \int_{k_{\rm min}}^{\Lambda/2} \frac{{\rm d}k}{2\pi} 
\Big(E(k) + \sum_{\ell=1}^{\ell_{\rm max}}\frac{S_{\ell}^2+P_{\ell}^2}{E(k)} 
\nonumber \\
& &  - \sqrt{k^2+1} \,\Big)
\label{d17} \\
& = & - \frac{L}{2\pi} \left( m^2-1+2\sum_{\ell=1}^{\ell_{\rm max}}(S_{\ell}^2+P_{\ell}^2)\right)\ln \Lambda + \Delta \tilde{E}_3. 
\nonumber
\end{eqnarray}
$\Delta \tilde{E}_3$ is the finite part which can again be computed analytically.
Finally, $\Delta E_4$ is the (vacuum subtracted) double counting correction, 
\begin{eqnarray}
\Delta E_4 &=& \frac{L}{2\pi} \left( m^2-1+2\sum_{\ell=1}^{\ell_{\rm max}} (S_{\ell}^2+P_{\ell}^2)\right)(\gamma + \ln \Lambda)
\nonumber \\
& & -\frac{\gamma L }{\pi}(m-1).
\label{d18}
\end{eqnarray}
In the sum $\Delta E_3+\Delta E_4$ the $\ln \Lambda$ terms are cancelled exactly and a finite result for $M_{\rm B}/N$, Eq.~(\ref{d13}),
is obtained.
\subsection{Finding the self-consistent potential}\label{sect4d}
In non-relativistic HF calculations, one usually determines the self-consistent potential iteratively. Starting from some guess,
one solves the HF equation, computes the new HF potential and repeats this procedure until it has converged. This method was also 
employed
in Ref.~\cite{L9} for the 't~Hooft model, using a lattice discretization in coordinate space. We do not have this option here.
The reason is the fact that we work in the continuum and have already sent the UV cutoff to $\infty$ and the bare 
coupling constant to 0. As is clear from the self-consistency relations (\ref{a3}), in this case the HF potential is
determined by the far UV region where lowest order perturbation theory holds. Therefore we always get back the potential which we put in,
irrespective of what we choose. If we would keep the cutoff finite like on the lattice, we could in principle use an iteration, but the
convergence would 
slow down with increasing cutoff. A way out of this problem was found in Ref.~\cite{L20} for the GN model where   
the HF energy of the baryon was simply minimized with respect to the potential $S(x)$.
Technically, the minimization was done with respect to the Fourier coefficients $S_{\ell}$ of $S(x)$.
In the NJL$_2$ case there are twice as many
parameters due to the pseudoscalar potential, so that this method would be rather cumbersome. Moreover,
owing to the analytical calculations in Secs.~\ref{sect2} and \ref{sect3}, we already have a fairly good qualitative idea of
 how the potentials should look like.
Hence the following strategy has turned out to be more economic. We
write down a parametrization of $S$ and $P$ which is both flexible and well suited to the known limiting cases.
We then vary with respect to the parameters. Since we are only interested in a limited precision, a relatively small number of parameters
is sufficient, provided the parametrization is judiciously chosen. The convergence can be checked by comparing runs with different 
numbers of parameters. A natural ansatz for trial functions in the present problem is given by 
\begin{equation}
S(x) = \sum_{k=1}^K \frac{c_k}{\cosh^{2k}\xi}, \quad
P(x)  =  \sum_{k=1}^K \frac{d_k  \sinh \xi}{\cosh^{2k+1}\xi}
\label{d19}
\end{equation}
($\xi=yx)$. This is clearly compatible with the no-sea effective theory, see Eq.~(\ref{c8}). The derivative expansion
would suggest even powers of $1/\cosh \xi $
in $P(x)$ at low $\gamma$ rather than odd powers, see Eq.~(\ref{b4}), but this makes almost no difference in practice, once
we keep a sufficient number
of terms. The coefficients $c_k,d_k$ and the scale parameter $y$ serve as variational parameters. The baryon mass computed as 
explained above is a function of these parameters and is minimized in the ($2K+1$)-dimensional parameter space via some standard algorithm.
\subsection{Numerical results}\label{sect4e}
All the results shown below were obtained by keeping the first three terms in each sum, Eq.~(\ref{d19}), and varying
with respect to 6 coefficients
($c_k,d_k$) and the scale parameter $y$. We found that a matrix dimension $804 \times 804$ (corresponding to $\bar{N}=200$ in
 Sec.~\ref{sect4a})
and box size $L=20$ (in units where $m=1$)
were adequate. The minimum in the 7-dimensional space was found with a standard conjugate gradient method involving 70 iteration
steps. We first tested the whole procedure for several values of $\nu,\gamma$ where the analytical approaches are expected to be reliable.
For example, at $\gamma=0.2,\nu=1$ the derivative expansion gives $M_{\rm B}=0.52227$ and the numerical HF 
calculation $M_{\rm B}=0.52248$.
This is satisfactory, especially since the numerical value is the result of subtracting large numbers 
(of the order of 10 000) due to the Dirac sea, cf. Eq.~(\ref{d14}).
The scalar and pseudoscalar HF potentials are indistinguishable on a plot if one compares the derivative expansion with the 
full HF calculation, therefore we don't show them here.
At $\gamma=5.0,\nu=1$ the no-sea effective theory
yields $M_{\rm B}=0.98847$ as compared to the numerical HF calculation $M_{\rm B}=0.98850$, again with excellent agreement
of the self-consistent potentials. We also tested the HF calculation against the no-sea effective theory at small filling.
For $\gamma=0.2,\nu=0.2$ the analytical value $M_{\rm B}=0.197314$
compares well with the numerical result $M_{\rm B}=0.197337$. These examples give strong support to the numerical method
as well as to the (independent) analytical calculations.

\begin{figure}
\begin{center}
\epsfig{file=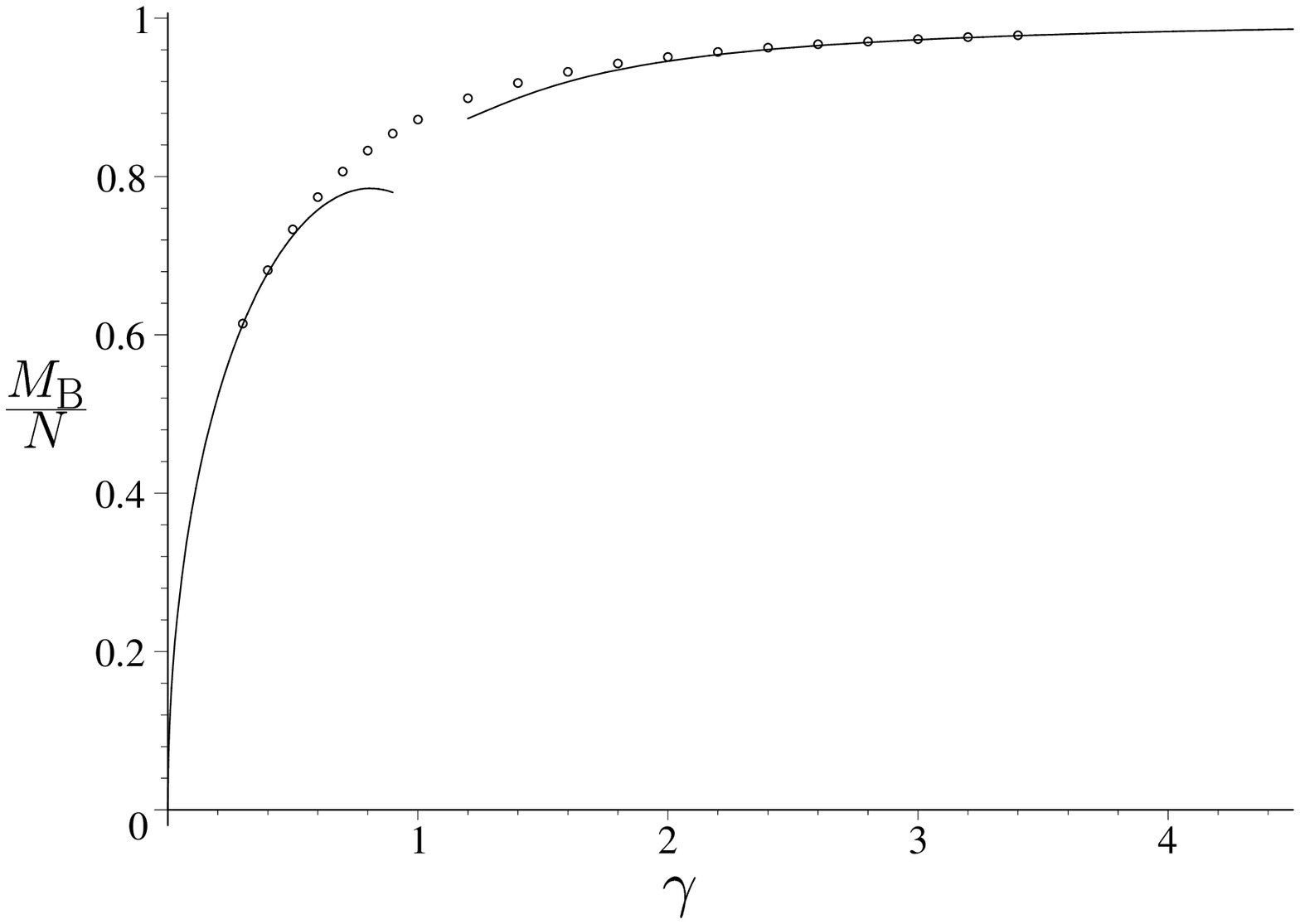,angle=270,width=8cm}
\caption{Circles: numerical results for baryon mass at $\nu=1$ versus $\gamma$. Curves: analytical asymptotic
predictions (derivative expansion to the left, no-sea effective theory to the right).}
\label{fig4}
\end{center}
\end{figure}
\begin{figure}
\begin{center}
\epsfig{file=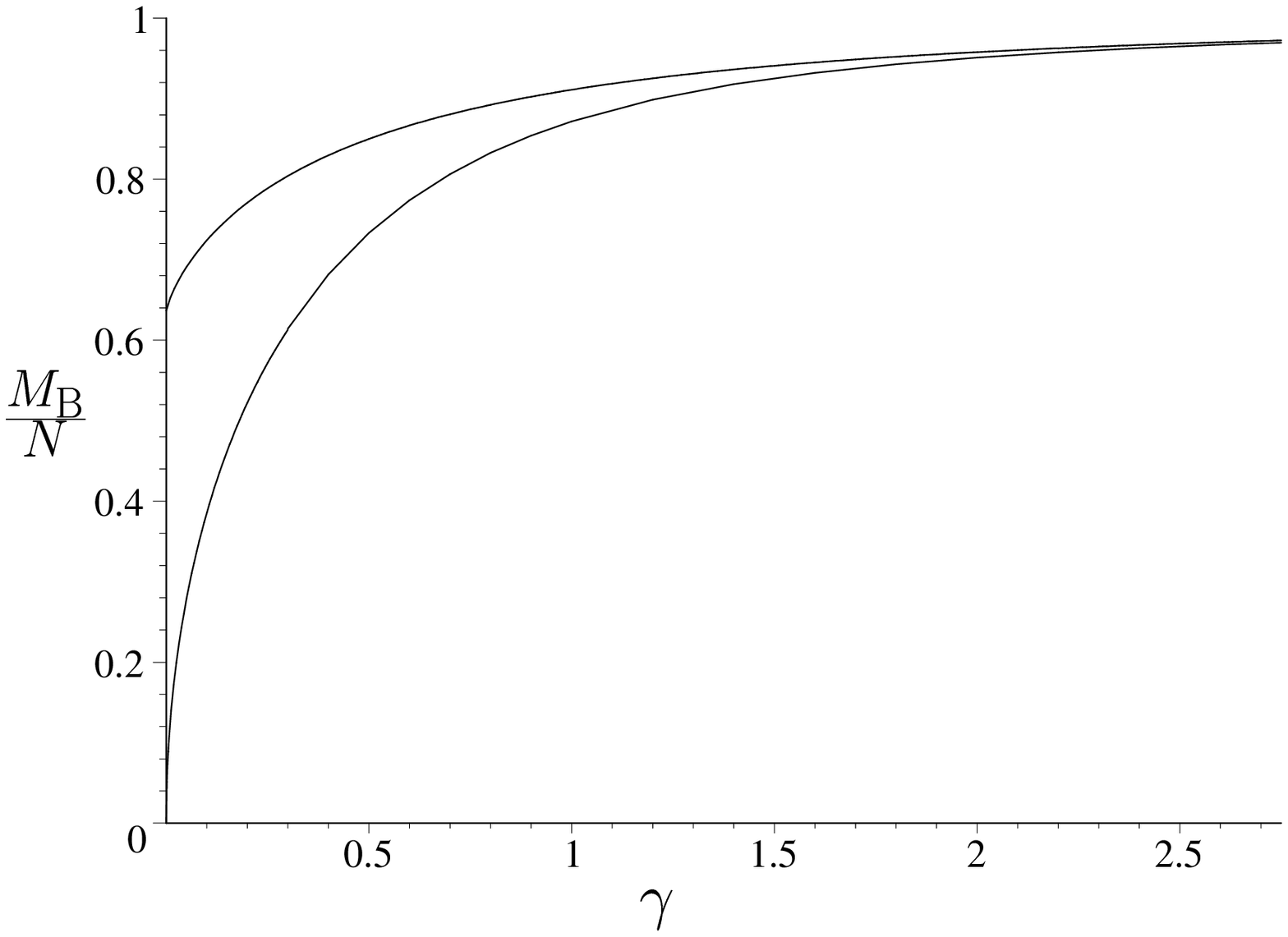,angle=270,width=8cm}
\caption{Comparison of baryon mass for large $N$ GN (upper curve) and NJL$_2$ (lower curve) models, at $\nu=1$.
These curves can be regarded equally well as zero temperature phase boundaries in the ($\mu,\gamma$)-plane.}
\label{fig5}
\end{center}
\end{figure}
\begin{figure}
\begin{center}
\epsfig{file=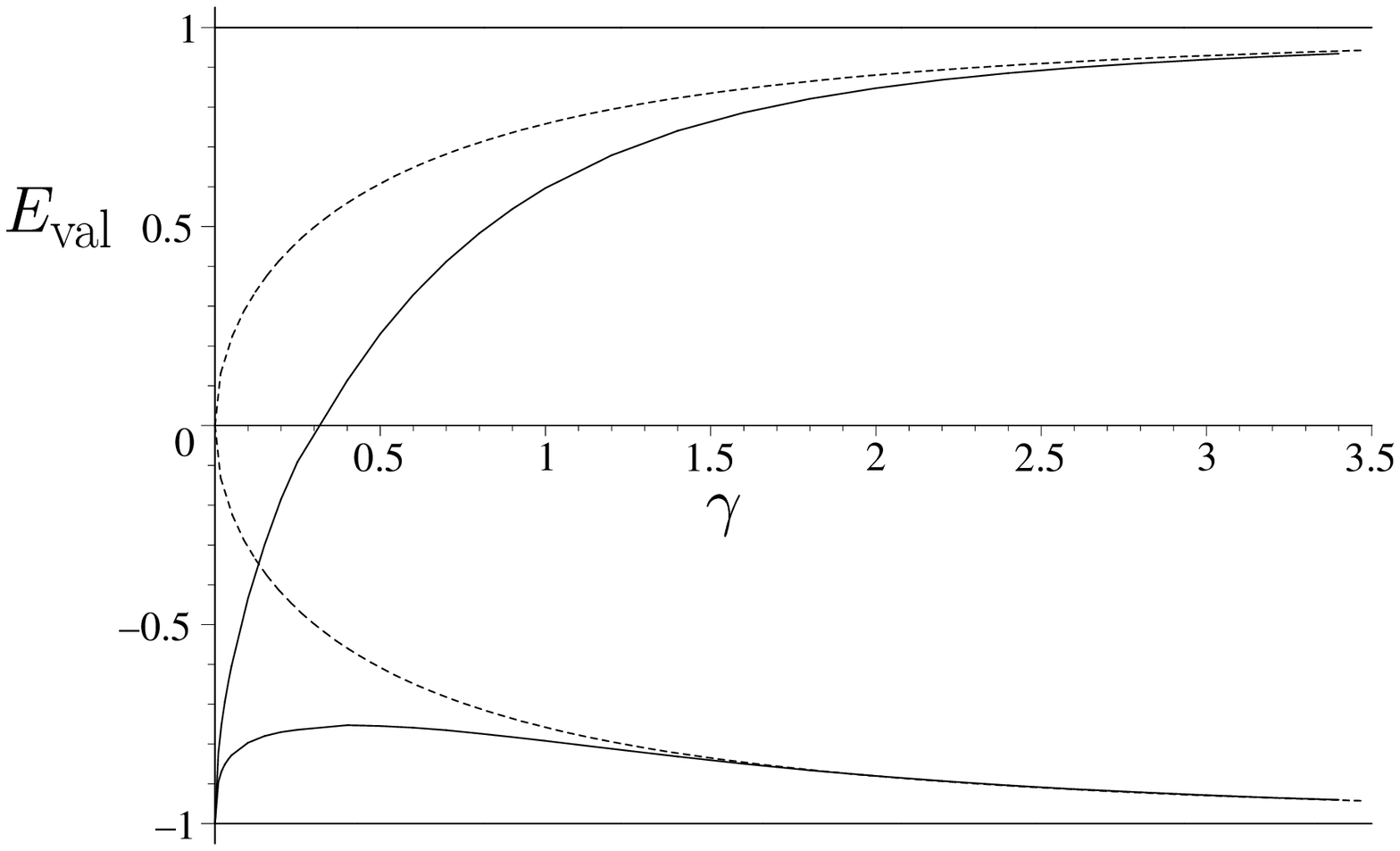,angle=270,width=8cm}
\caption{Energies of valence levels for GN model (dashed curves) and NJL$_2$ model (solid curves) versus $\gamma$, illustrating
how the NJL$_2$ baryon interpolates between a Skyrme-type baryon with induced fermion number and a conventional,
non-relativistic, valence-type baryon.}
\label{fig6}
\end{center}
\end{figure}
Regarding the phase diagram, the question of most immediate interest is the dependence of $M_{\rm B}$ on $\gamma$ for
full occupation $\nu=1$.
This is expected to yield a critical (2nd order) curve in the phase diagram at $T=0$. In Fig.~\ref{fig4} we show the numerical points together
with the asymptotic
expansions, using both derivative expansion and no-sea effective theory. In Fig.~\ref{fig5} we compare this newly calculated 
baryon mass in the massive NJL$_2$
model with the known one from the massive GN model, choosing a somewhat smaller range of $\gamma$ as compared to Fig.~\ref{fig4} 
to highlight the differences. The two curves shown reflect directly the critical lines in the ($\mu,\gamma$)-plane at $T=0$ of the 
respective phase diagrams. 
As a by-product of the calculation of baryon masses we also get the single particle energy of the valence
levels, i.e., discrete levels inside the mass gap. It is instructive to compare these energies between the GN and NJL$_2$ models as well.
Fig.~\ref{fig6} shows the pair of (charge conjugation) symmetric states in the GN model which come together at zero energy at $\gamma=0$
(this point corresponds to a kink and antikink at infinite separation). 
The baryons in the NJL$_2$ model also feature a pair of
discrete states which approach those of the GN model at large $\gamma$. At small $\gamma$, they converge to the
lower edge of the mass gap, the upper level crossing zero near $\gamma_0=0.3$. Both of these valence levels are fully
occupied. Below $\gamma_0$ one would talk about induced fermion number (the number of negative energy levels
changes as compared to the vacuum), whereas above $\gamma_0$ one has ordinary valence fermions like in the 
GN model. Note however that nothing discontinuous happens at the ``spectral flow" point $\gamma_0$. This picture nicely
illustrates the transition
from the massless baryon at $\gamma=0$ (with baryon number given by winding number of the pion field, like in the Skyrmion case)
to the weakly bound non-relativistic baryon emerging at large $\gamma$ in either GN or NJL$_2$ models. The latter one
is apparently insensitive to the original type of chiral symmetry, be it U(1) or Z$_2$.

\begin{figure}
\begin{center}
\epsfig{file=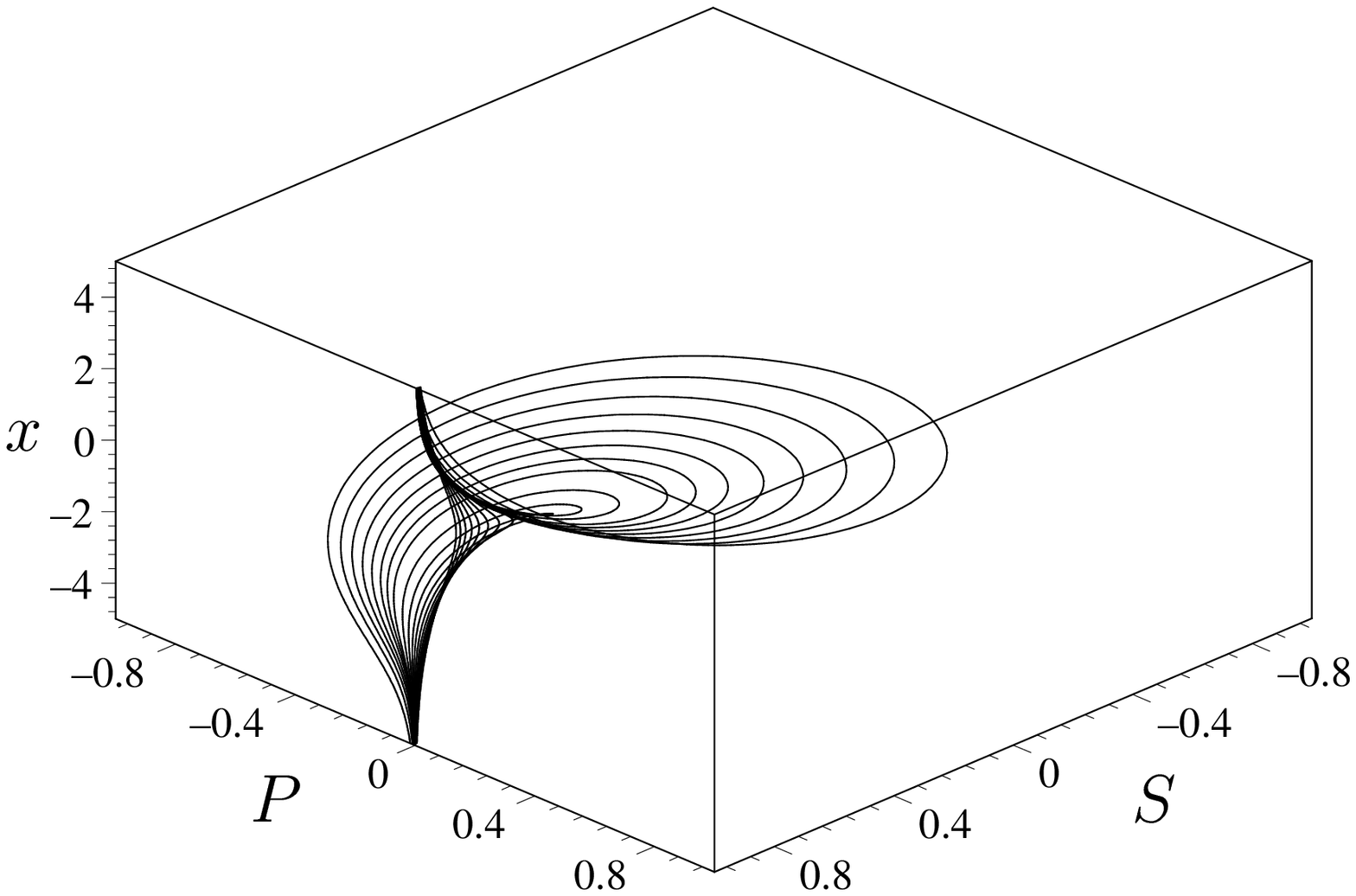,height=7cm,width=7cm,angle=270}
\caption{HF potentials $S,P$ versus $x$ in a 3d plot. The curves correspond to $\gamma=0.3,...,1.0$
in steps of 0.1 and $\gamma=1.2,...,3.4$ in steps of 0.2. The outermost curve has the smallest value of $\gamma$. 
The curves at large $\gamma$ accumulate around $S=1,P=0$ and are not well resolved, see also Figs.~\ref{fig8}--\ref{fig10}}
\label{fig7}
\end{center}
\end{figure}
\begin{figure}
\begin{center}
\epsfig{file=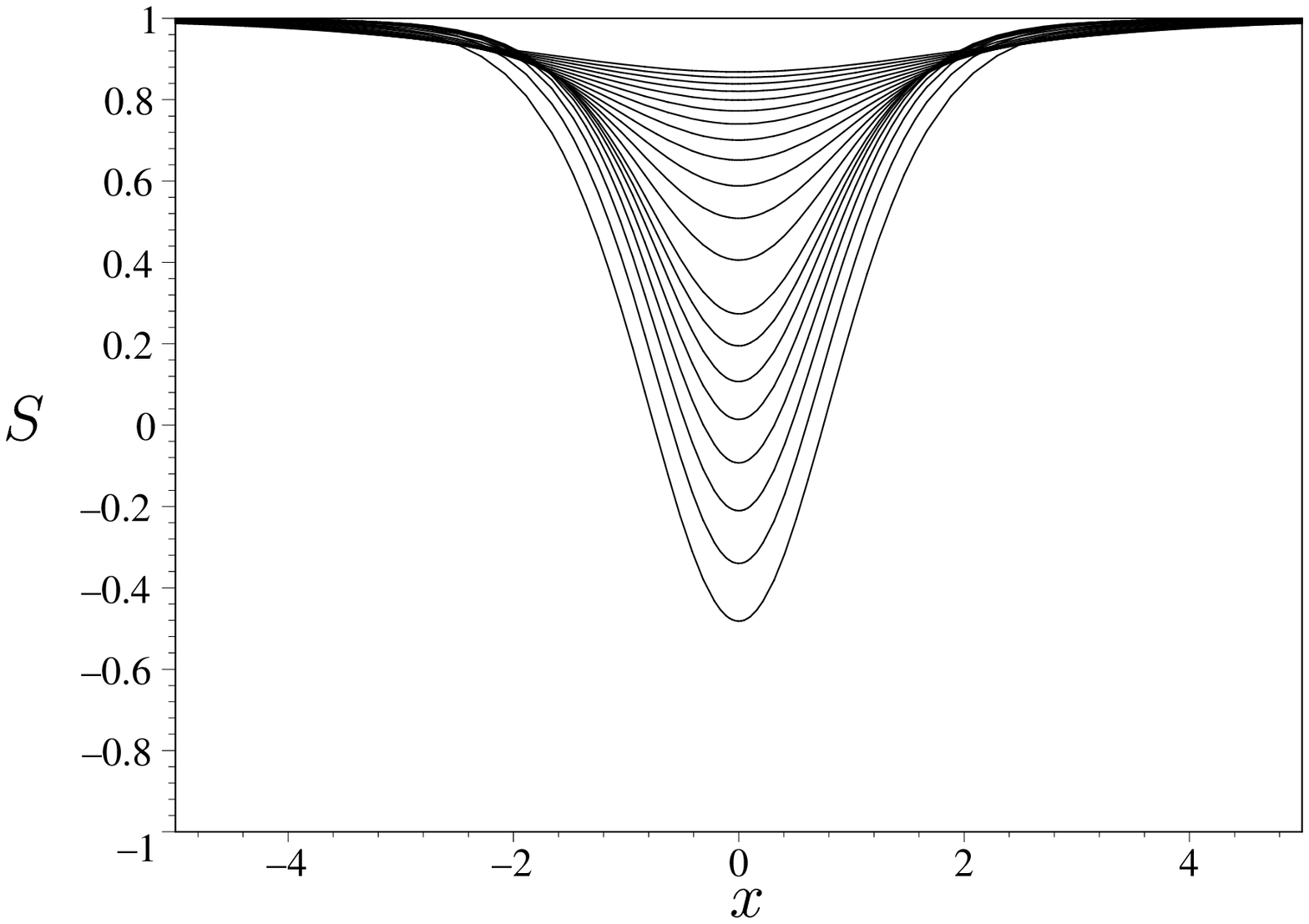,angle=270,width=8cm}
\caption{Projection of Fig.~\ref{fig7} onto the ($x,S$)-plane, showing how the shape of the scalar potential
evolves with $\gamma$.}
\label{fig8}
\end{center}
\end{figure}
\begin{figure}
\begin{center}
\epsfig{file=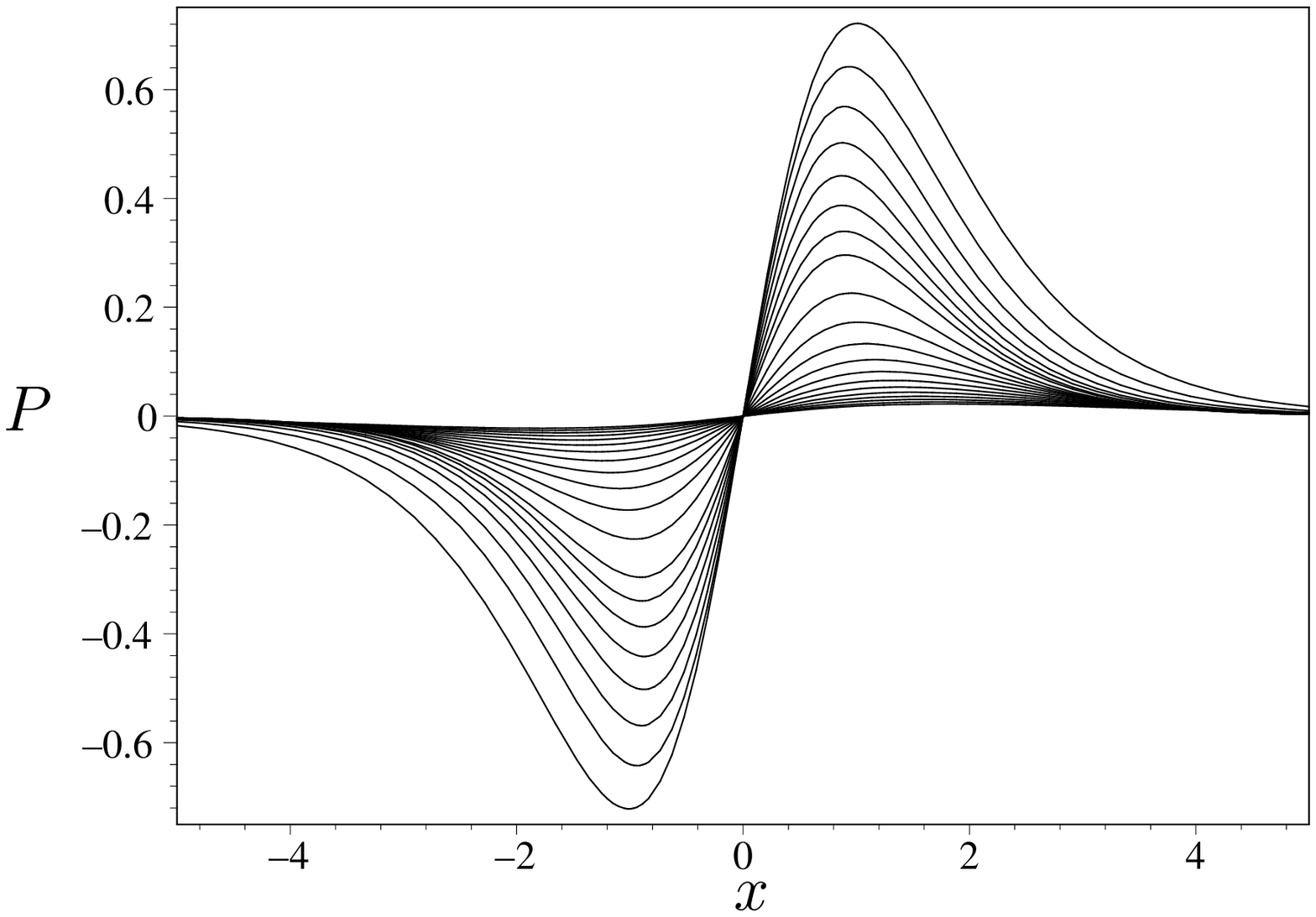,angle=270,width=8cm}
\caption{Analogous to Fig.~\ref{fig8}, but for the ($x,P$)-plane and pseudoscalar potential.}
\label{fig9}
\end{center}
\end{figure}
\begin{figure}
\begin{center}
\epsfig{file=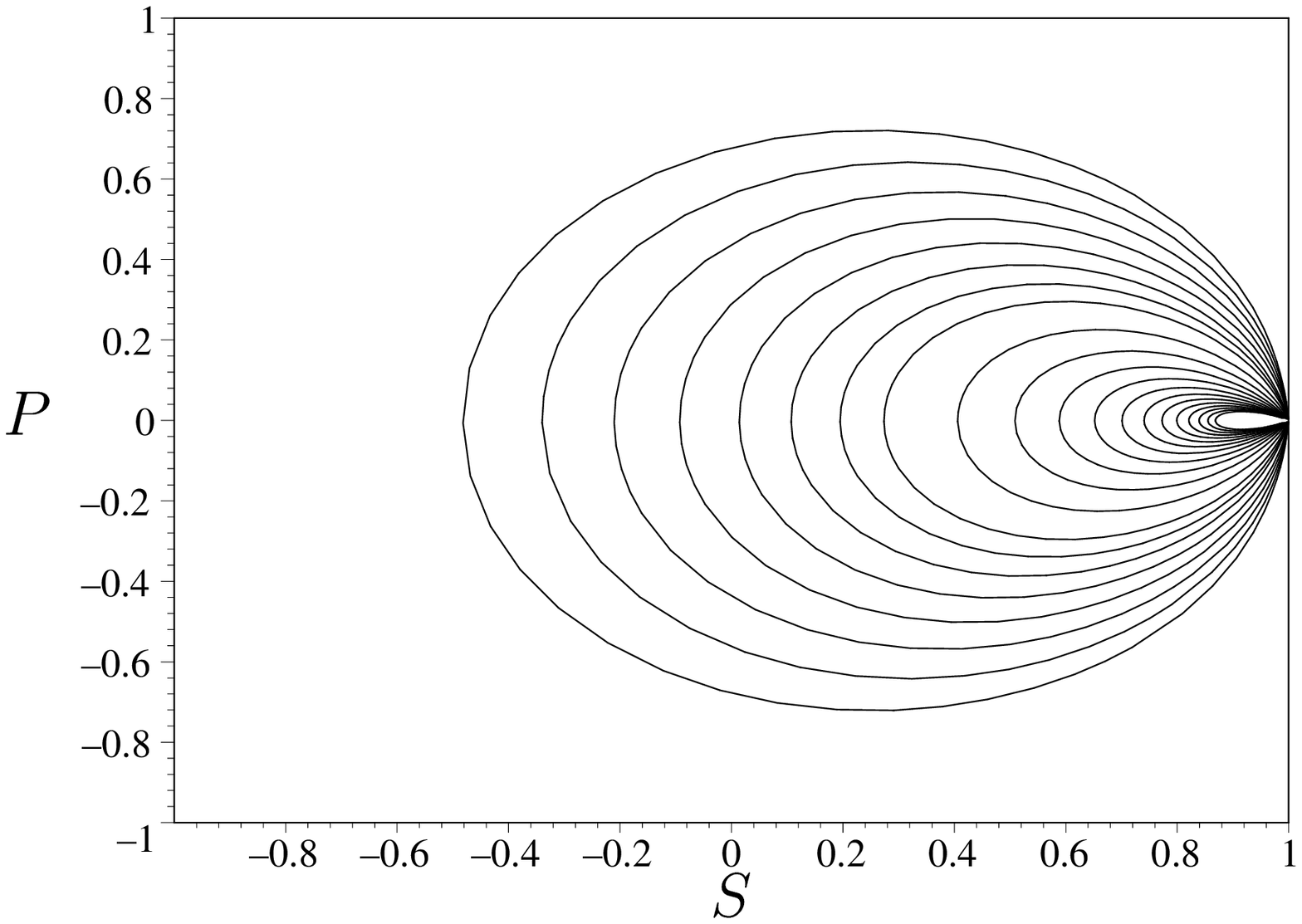,height=7cm,width=7cm,angle=270}
\caption{Projection of Fig.~\ref{fig7} onto the ($S,P$)-plane.}
\label{fig10}
\end{center}
\end{figure}
The HF potentials belonging to these calculations are shown in Figs.~\ref{fig7}-\ref{fig10}. Fig.~\ref{fig7}
 is a 3d-plot containing the full information 
how $S$ and $P$ depend on $x$ for 20 different values of $\gamma$. 
The 3 different projections of this plot onto the 
coordinate planes show the $x$-dependence of $S$ (Fig.~\ref{fig8}) and $P$ (Fig.~\ref{fig9}) in more detail, as well as 
parametric contour plots
in the $(S,P)$-plane (Fig.~\ref{fig10}). We can infer from this last figure that the chiral winding number jumps from 1 to 0 slightly below
$\gamma=0.7$, where the contour plot hits the origin (close to the 5th curve from the outside). Note that this does not
coincide with the value $\gamma_0=0.3$ where the upper valence level crosses zero, so that induced fermion number
is not directly related to winding number of $\Phi=S-{\rm i}P$ for larger bare fermion masses. This is in agreement
with the general observation that the integer part of induced fermion number is not topological, being sensitive
to spectral flow \cite{L21}.

\begin{figure}
\begin{center}
\epsfig{file=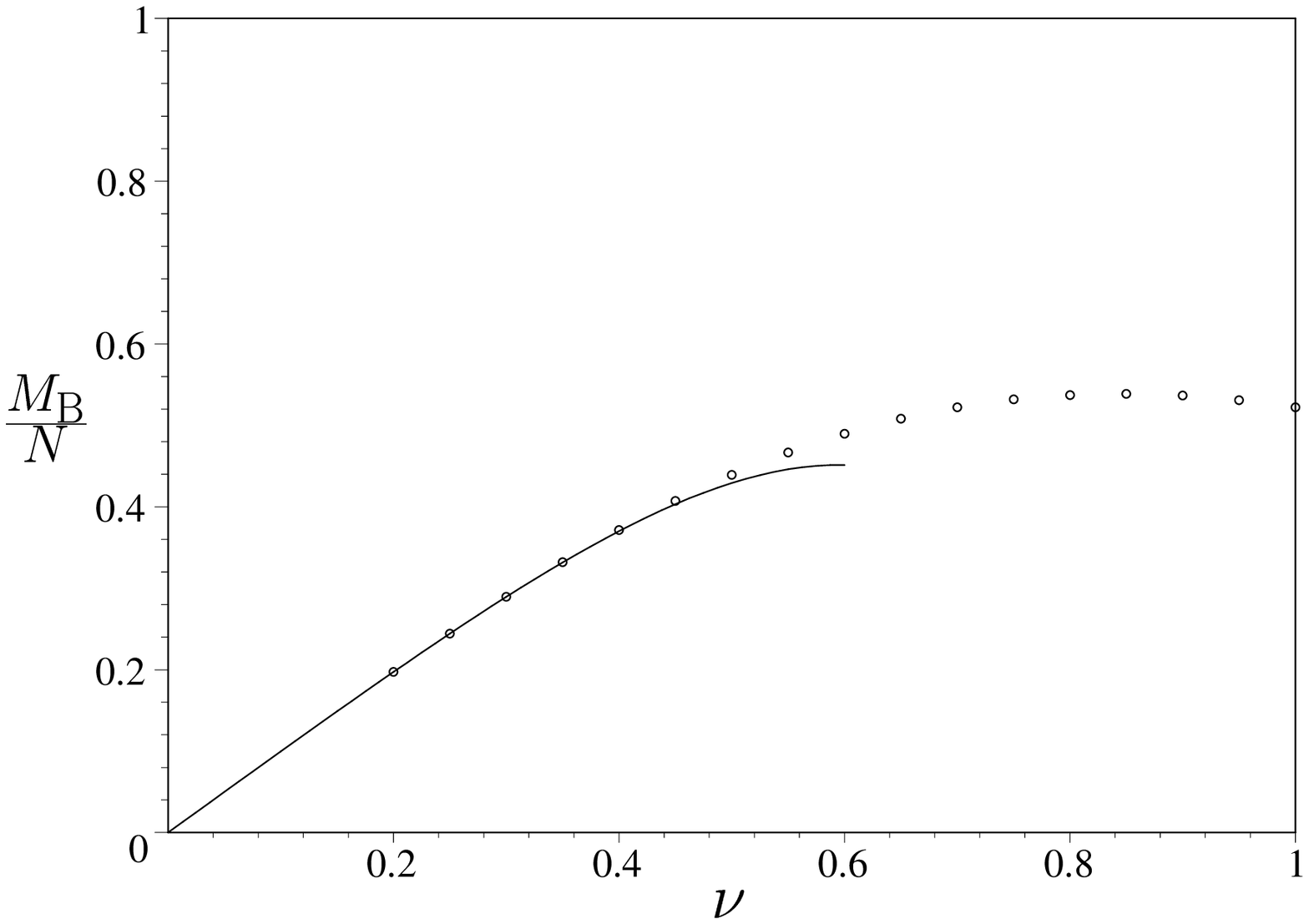,angle=270,width=8cm}
\caption{Baryon mass in the NJL$_2$ model versus filling fraction $\nu$ of upper valence level, for $\gamma=0.2$. Circles: numerical
HF calculation, curve: analytical prediction based on the no-sea effective theory.}
\label{fig11}
\end{center}
\end{figure}
\begin{figure}
\begin{center}
\epsfig{file=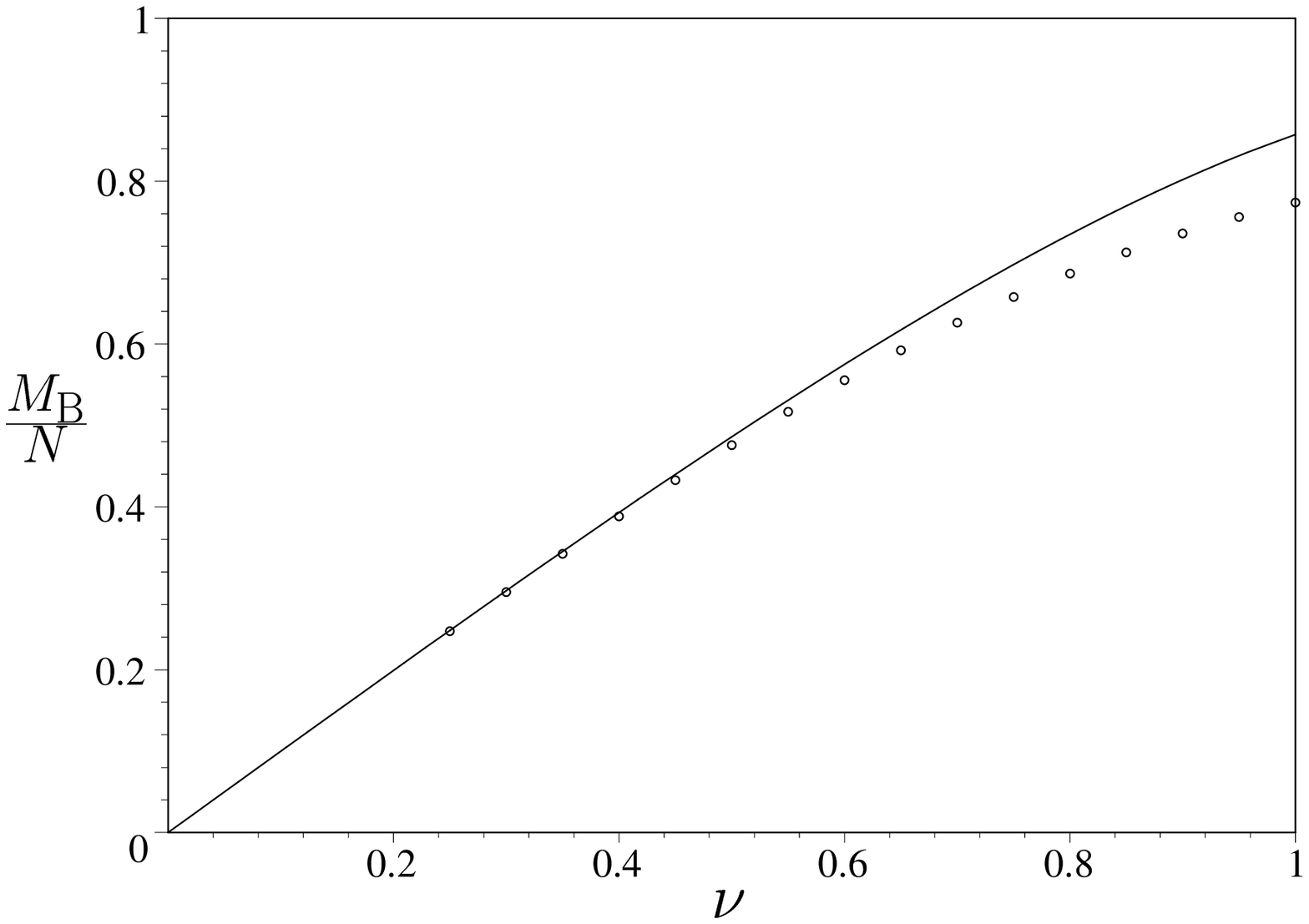,angle=270,width=8cm}
\caption{Same as Fig.~\ref{fig11}, but for $\gamma=0.6$.}
\label{fig12}
\end{center}
\end{figure}
\begin{figure}
\begin{center}
\epsfig{file=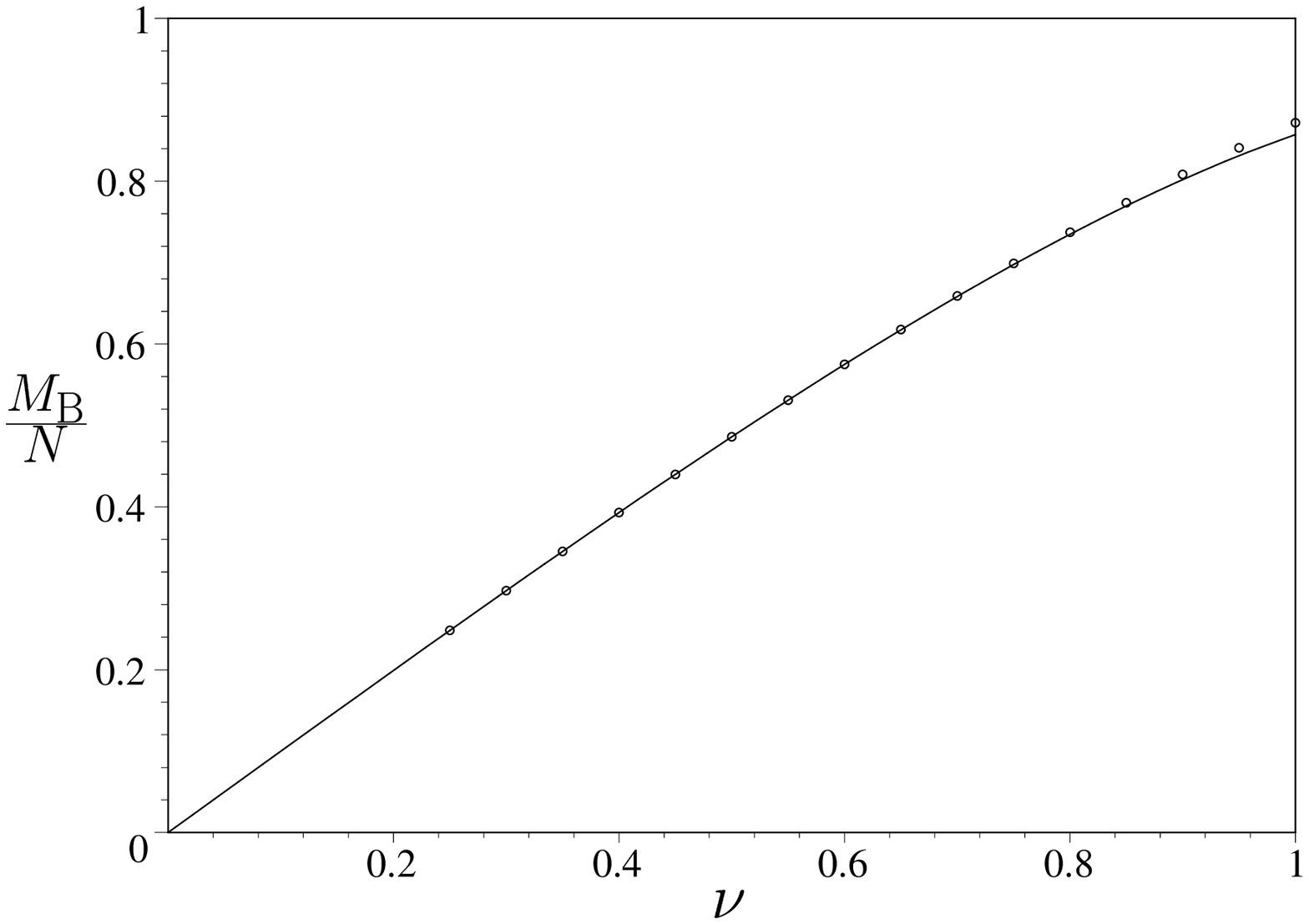,angle=270,width=8cm}
\caption{Same as Fig.~\ref{fig11}, but for $\gamma=1.0$.}
\label{fig13}
\end{center}
\end{figure}
Although it is presumably not relevant for the phase diagram, we now turn to baryons with partially occupied valence level.
The HF calculation can be done exactly as before, except that the single particle energy of the upper valence level is weighted by
the occupation fraction $\nu$ when calculating the baryon mass.
We have carried out detailed studies of the $\nu$-dependence for the three values $\gamma =0.2,0.6,1.0$  
and compared the HF results with the no-sea effective theory expected to describe correctly the $\nu \to 0$ behavior.
The results for the baryon mass are shown in Figs.~\ref{fig11}-\ref{fig13} and again underline the value of the much simpler analytical
results. 

\begin{figure}
\begin{center}
\epsfig{file=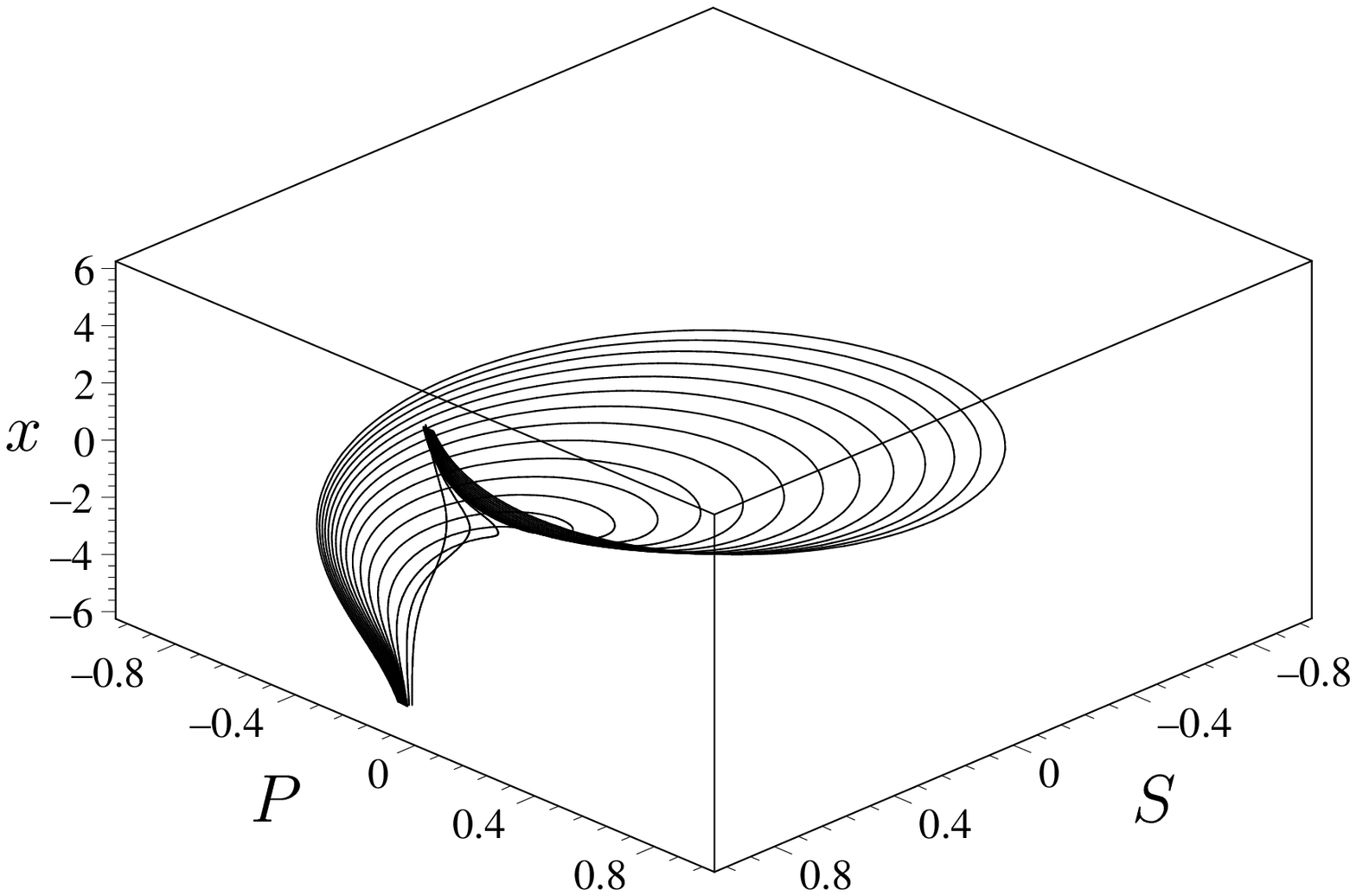,height=7cm,width=7cm,angle=270}
\caption{HF potentials $S,P$ versus $x$ in a 3d plot. The 17 curves correspond to $\nu=0.2,...,1.0$
in steps of 0.05. The outermost curve has the largest value of $\nu$.}
\label{fig14}
\end{center}
\end{figure}
The self-consistent potentials as a function of $\nu$ are illustrated for the case $\gamma=0.2$ in Fig.~\ref{fig14}
(the other cases are qualitatively similar). This 3d-plot corresponds to Fig.~\ref{fig7}, except that now the different curves are obtained
by varying $\nu$ at fixed $\gamma$ rather than $\gamma$ at fixed $\nu$. The similarity between Figs.~\ref{fig7} and \ref{fig14}
is quite remarkable.
Depleting the valence level has apparently a very similar effect as increasing the bare fermion mass, i.e., it drives the NJL$_2$
baryon into the non-relativistic regime where it becomes indistinguishable from the GN baryon. 

\begin{figure}
\begin{center}
\epsfig{file=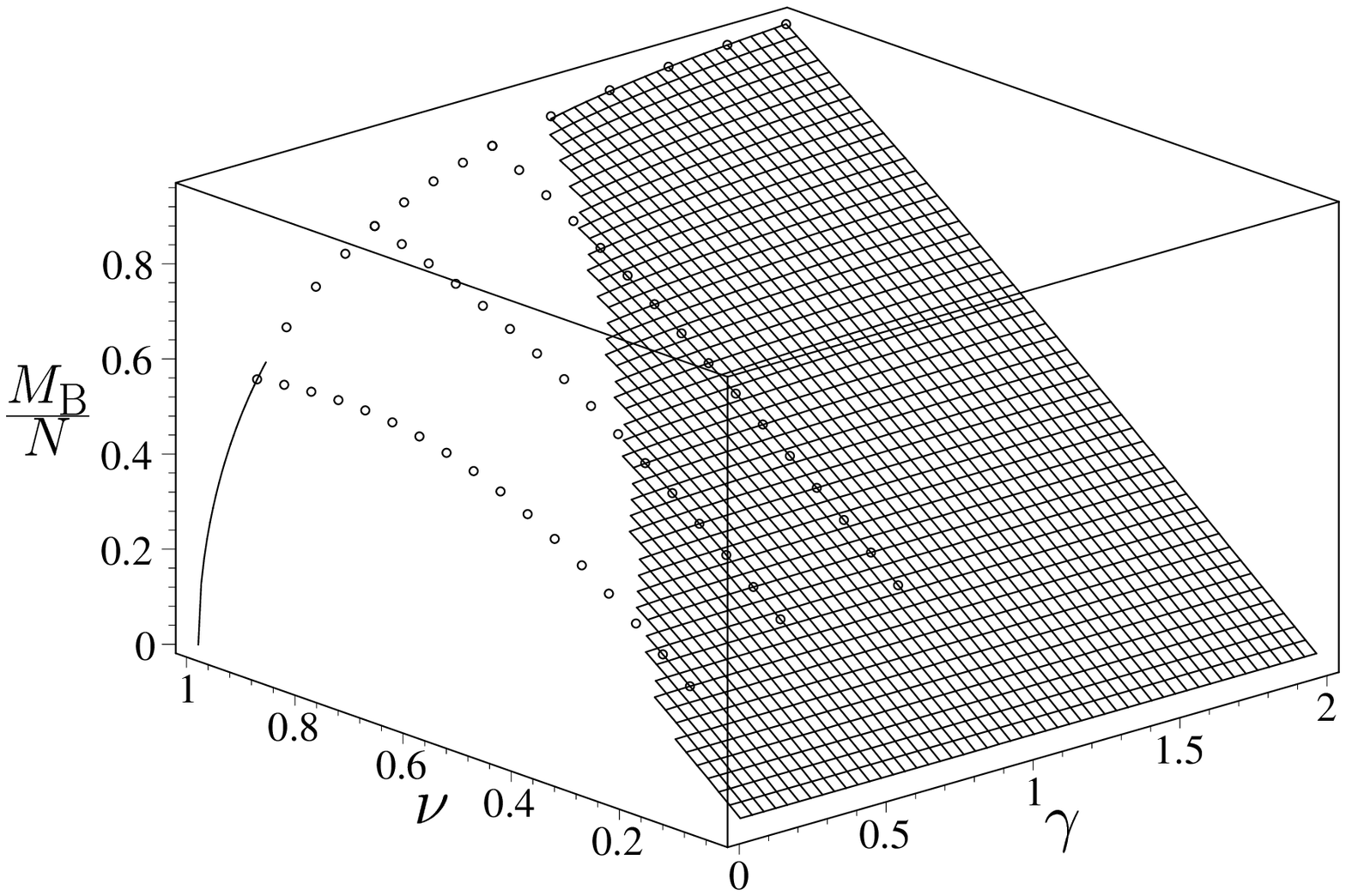,height=8cm,width=6.4cm,angle=270}
\caption{Baryon mass as a function of filling fraction $\nu$ and confinement parameter $\gamma$. The numerical
HF results (circles) have been combined with the analytical results of Fig.~\ref{fig3}, completing the picture of what is known
by now.}
\label{fig15}
\end{center}
\end{figure}
In order to summarize the results obtained in the present work, we have combined all the available points from the numerical HF calculations
with the analytical results taken from Fig.~\ref{fig3}, see Fig.~\ref{fig15}. The numerical points start to fill the white gap in Fig.~\ref{fig3} 
and match perfectly onto 
the analytical predictions in the region where these can be trusted. This can be seen more clearly in the 2d-plots of Figs.~\ref{fig4}
and Figs.~\ref{fig11}--\ref{fig13}. We do not show any results for $\gamma>2$ where the asymptotic formula (\ref{c20}) is fully adequate.
Notice that when the present work was started, only the short solid curve starting at $\nu=1,\gamma=0$ had been known
from the derivative expansion \cite{L10}.

\section{Summary and conclusions}\label{sect5}
This paper is part of an ongoing effort to establish the phase diagram of (large $N$) Gross-Neveu type models.
Ultimately, we would like to construct the phase diagram of the massive NJL$_2$ model in $(\gamma,\mu,T)$
space in a similarly complete fashion as what has already been achieved for the massive GN model.
One important building block which was still missing are the baryons in the massive NJL$_2$ model. The
only source of information so far was the derivative expansion, restricted to the vicinity of the chiral limit
and to completely filled single particle levels. After briefly reviewing these results, we have identified another 
parameter region where systematic analytical approximations can be performed, namely the non-relativistic 
regime (heavy fermions or weak filling of the valence level). Here, a recently developed no-sea effective theory
has proven to be  very efficient. The gap between these two asymptotic approaches could only be filled at the 
expense of full numerical HF calculations
including the Dirac sea. As a result, we have obtained a rather comprehensive picture
of how the baryon evolves from a Skyrme-type topological object in the chiral limit to a non-relativistic valence bound state.
The transition from induced fermion number to valence fermion number has a simple 
interpretation in terms of spectral flow as a function of the confinement parameter, as discussed in connection 
with Fig.~\ref{fig6}. In the heavy fermion limit, it becomes irrelevant whether the model had originally a U(1) or Z$_2$
chiral symmetry, and the results for the massive NJL$_2$ and GN models converge.

As compared to the massive GN model with its complete analytical solution, the analysis of the massive NJL$_2$ model
is significantly more involved. Although the self-consistent potentials which we find numerically appear to have simple
shapes, we have not been able to come up with an analytical solution. One source of difficulty 
is the fact that the potentials in the NJL$_2$ model are not reflectionless. Related to this,
even in those limits where we have analytical control, the potentials do not lead to any known
exactly solvable Dirac equation. If this problem is analytically tractable at all,
the techniques must be quite different from those which have been successful in the GN model case.

Finally, coming back to the phase diagram, we have in fact determined one particular phase boundary in the present work.
Like in the GN model, the $T=0$ base line of the phase diagram in ($\gamma,\mu,T$) space is expected to be given by the curve
$M_{\rm B}(\gamma)$ of Figs.~\ref{fig4} and \ref{fig5}. Another piece of information about the phase diagram is the vicinity of the tricritical
point discussed recently in Ref.~\cite{L22}. To complete this picture remains quite a challenge.

\newpage 
\appendix
\section{Results for coefficients defined in Sec.~\ref{sect3}}
Here we collect the $\gamma$-dependent coefficients which enter the results of the no-sea effective theory
in Sec.~III. Coefficients for the scalar potential $S$, Eq.~(\ref{c8}),
\begin{eqnarray}
s_{22} & = & - \frac{1}{4(1+ \gamma)^2}, \quad
s_{42} \ = \ \frac{5 \gamma^2+ 3\gamma +2}{96\gamma(1+\gamma)^5}
\nonumber \\
s_{44} & = & - \frac{3\gamma+4}{64 \gamma(1+\gamma)^4}
\nonumber \\
s_{62} & = & - \frac{91 \gamma^4 -102 \gamma^3+275 \gamma^2 + 276 \gamma+88}{23040 \gamma^2 (1+\gamma)^8}
\nonumber \\
s_{64} & = & \frac{81 \gamma^3+13 \gamma^2-140 \gamma-24}{9216 \gamma^2(1+\gamma)^7}
\nonumber \\
s_{66} & = & - \frac{45 \gamma^2-136 \gamma-60}{9216 \gamma^2 (1+\gamma)^6}
\label{c9}
\end{eqnarray}
Coefficients for the pseudoscalar potential $P$, Eq.~(\ref{c8}),
\begin{eqnarray}
p_{33} & = & \frac{1}{8 \gamma(1+\gamma)^2}, \quad
p_{53} \ = \ - \frac{(3\gamma +2)(\gamma-3)}{192 \gamma^2(1+\gamma)^5}
\nonumber \\
p_{55} & = & \frac{\gamma-2}{64\gamma^2 (1+\gamma)^4}
\label{c11}
\end{eqnarray}
Coefficients for the valence fermion density $\rho_{\rm val}$, Eq.~(\ref{c17}),
\begin{eqnarray}
v_{22} & = & \frac{1}{4(1+\gamma)}, \quad
v_{42} \ = \ - \frac{\gamma^2+ \gamma +1}{48 \gamma (1+\gamma)^4}
\nonumber \\
v_{44} & = &  \frac{\gamma + 4}{64 \gamma (1+\gamma)^3}
\nonumber \\
v_{62} & = &  \frac{2\gamma^4+10 \gamma^3+49\gamma^2+37\gamma+11}{2880\gamma^2 (1+\gamma)^7}
\nonumber \\
v_{64} & = & - \frac{15\gamma^3+151\gamma^2+76\gamma+48}{9216 \gamma^2(1+\gamma)^6}
\nonumber \\
v_{66} & = & \frac{9\gamma^2-40\gamma+12}{9216\gamma^2(1+\gamma)^5}
\label{x1}
\end{eqnarray}
Coefficients for the induced fermion density $\rho_{\rm ind}$, Eq.~(\ref{c18}),
\begin{eqnarray}
i_{42} & = & \frac{1}{16 \gamma (1+\gamma)^3}, \quad
i_{44} \ = \ - \frac{3}{32 \gamma (1+\gamma)^3}
\nonumber \\
i_{62} & = & - \frac{2\gamma^2 - 5\gamma-4}{192 \gamma^2 (1+\gamma)^6}, \quad
i_{64} \ = \  \frac{4 \gamma^2-7\gamma-8}{128 \gamma^2(1+\gamma)^6}
\nonumber \\
i_{66} & = & - \frac{5 (\gamma-2)}{256 \gamma^2 (1+\gamma)^5}
\label{x2}
\end{eqnarray}

\newpage

\end{document}